\documentclass[aps,pre,preprint,groupedaddress]{revtex4-1}

\usepackage{graphicx}
\usepackage{grffile}
\usepackage{amsmath}

\begin{document}


\title{A fast algorithm for a three-dimensional synthetic model of intermittent turbulence}



\author{Francesco Malara}
\email[]{francesco.malara@fis.unical.it}
\author{Francesca Di Mare}
\author{Giuseppina Nigro}
\affiliation{Dipartimento di Fisica, Universit\`a della Calabria, Ponte P. Bucci, cubo 31 C, 87036 Rende (CS), Italy}

\author{Luca Sorriso-Valvo}
\affiliation{Nanotec/CNR, U.O.S. di Rende, Ponte P. Bucci, cubo 31 C, 87036 Rende (CS), Italy}


\date{\today}

\begin{abstract}
Synthetic turbulence models are a useful tool that provide realistic representations of turbulence, necessary to test theoretical results, to serve as background fields in some numerical simulations, and to test analysis tools. Models of 1D and 3D synthetic turbulence previously developed still required large computational resources. A new ``wavelet-based" model of synthetic turbulence, able to produce a field with tunable spectral law, intermittency and anisotropy, is presented here. The rapid algorithm introduced, based on the classic $p$-model of intermittent turbulence, allows to reach a broad spectral range using a modest computational effort. 
The model has been tested against the standard diagnostics for intermittent turbulence, i.e. the spectral analysis, the scale-dependent statistics of the field increments, and the multifractal analysis, all showing an excellent response. 
\end{abstract}

\pacs{}
\keywords{Synthetic turbulence, intermittency}

\maketitle

\section{Introduction}

Turbulence represents an universal phenomenon characterizing the dynamics of different kinds of fluids, like gases, liquids, plasmas, etc., both in nature and in laboratory devices. It is responsible for the efficient transfer of energy across scales, making the connection between the macroscopic flow and the microscopic dissipation of its energy. Moreover, turbulence plays a key role in determining various phenomena. For instance, the anomalous diffusion of tracers in a flow may be controlled by the properties of turbulence, and the transport of charged particles in astrophysical or laboratory plasmas is determined by the properties of the turbulent magnetic field.

The phenomenological description of hydrodynamic turbulence by Kolmogorov~\cite{K41} consists of a superposition of velocity perturbations, whose energy is distributed over a wide range of spatial scales. Each scale is coupled with the other scales by nonlinear effects which allow for energy to be transferred from a given scale to another. Typically, it is assumed that energy is injected at large spatial scales (injection range) and is continuously transferred to smaller scales by nonlinear effects across the inertial range, finally reaching the smallest scales (dissipative range) where dissipation becomes effective. The energy transfer process (cascade) taking place in the inertial range gives rise to a typical power-law energy spectrum: $E(l)=E_0 (l/l_0)^\Gamma$, where $E(l)$ is the energy of fluctuations at the scale $l$ and $l_0$ is a given reference scale (typically, the integral scale) with $E_0=E(l_0)$. 
The exponent $\Gamma$ is determined by imposing the conservation of the spectral energy flux $\epsilon$ through the different scales of the inertial range: for an ordinary fluid it is $\Gamma=5/3$~\cite{K41}, but different values of $\Gamma$ can be considered in different contexts. For instance, in magnetohydrodinamics (MHD), assuming that nonlinear interactions are limited by the propagation of perturbations along the mean magnetic field, the value $\Gamma=3/2$ can be inferred~\cite{iro,kra}.

A peculiar property of turbulence is represented by intermittency~\cite{FRISCH}. Considering the increments $\Delta {\bf v}({\bf x},{\bf X})={\bf v}({\bf x}+{\bf X})-{\bf v}({\bf x})$ of the velocity field ${\bf v}$ at a given displacement ${\bf X}$ for all the possible positions ${\bf x}$, their statistical distribution $f(\Delta {\bf v})$ is not self-similar at all the scales $l=|{\bf X}|$. 
In particular, $f(\Delta {\bf v})$ is essentially Gaussian at large scales, while for decreasing $l$ the tails of the distribution $f(\Delta {\bf v})$ become more and more significant, indicating that fluctuations with amplitude much larger than the {\em rms} value become more and more frequent with decreasing the scale $l$. The lack of self-similarity reflects into a nonlinear dependence of the scaling exponents $\zeta_q$ of structure functions $S_q(l)$ on the order $q$, $S_q(l) \propto l^{\zeta_q}$ being the $q$-order moment of the distribution $f(\Delta {\bf v})$ at the scale $l$. 

Large-amplitude fluctuations at small scales appear to be localized in space. Thus, it has been speculated that intermittency is a consequence of a spatially nonuniform spectral energy flux. Fluctuating energy tends to concentrate at locations where the spectral flux is larger, and the energy localization becomes more noticeable at smaller scales due to a cumulative effect. Most of the models for the description of intermittent turbulence are based on this idea. Examples are the random-$\beta$ model~\cite{RANDOMBETA}, the She \& L\^eveque model~\cite{sheleveque}, and the $p$-model introduced by Meneveau \& Sreenivasan~\cite{meneveau87}. In the ``$p$-model", a 1D spatial distribution of the energy flux at different spatial scales is reconstructed through a multiplicative process. Thus, an eddy at a given scale $l$ breaks in two eddies at the scale $l/2$, and the energy flux $\epsilon$ associated with the parent eddy is unequally distributed to the two daughter eddies, with fractions given by $2p\epsilon$ and $2(1-p)\epsilon$ respectively, with $0.5 \le p \le 1$. Thus, going from larger to smaller scales the energy flux tends to become more and more spatially localized.

Intermittency is an intrinsic property of turbulence that has been found both in laboratory experiments~\cite{k62,anselmet} and in natural fluids like atmosphere and astrophysical plasmas~\cite{marsch&tu,horbury,sorriso99}. Thus, any model aimed at reproducing the main features of turbulence should include intermittency. A natural way to obtain a representation of turbulence is by direct simulations in which a numerical solution of fluid equations within a given spatial domain is calculated starting from suitable initial conditions. This approach has the advantage to be based on first principles (like mass, momentum and energy conservation); in particular, direct simulations of both fluid and MHD equations reproduce intermittency self-consistently. However, it is limited by finite space resolution which determines the extension of the range of spatial scales. Such a limitation can become very severe in 3D configurations: for high-Reynolds number fluids, as typically happens in astrophysical contexts, realistic simulations would require huge computational efforts. 

Another possibility to tackle a turbulence description is represented by ``synthetic turbulence". In this approach, the main properties of a turbulent field are reproduced starting from simplified models which mimic the processes taking place in real turbulence. The main advantage of synthetic turbulence is its reduced computational requirements with respect to direct simulations. This allows, for instance, to represent spatial scale ranges which are larger than in direct simulation, but employing smaller computing resources. Synthetic turbulence can be useful in particular contexts like: generating initial conditions for numerical simulations~\cite{rosales06}; describing processes which involve very different spatial scales (e.g., particle transport or acceleration, diffusion, drop formation)~\cite{sardina15}; understanding fundamental scaling properties of turbulence~\cite{juneja94,arneodo98}; evaluating sub-grid stresses~\cite{scotti99,kerstein01,mcdermott05}. Different methods have been proposed to generate synthetic turbulence with different features, according to the application they have been conceived for. Juneja et al.~\cite{juneja94} presented a ``wavelet-based'' model which produces a function with the statistical properties of a signal measured along a line in a turbulent field; in particular, intermittency is reproduced. 
A generalization in 3D of such a model has been proposed by Cametti et al.~\cite{cametti98}. 3D models obtained by a superposition of random-phased Fourier modes with a given spectrum have been used to study transport processes in turbulent magnetic fields (Zimbardo et al.~\cite{zimbardo00}, Ruffolo et al.~\cite{ruffolo06}); such models can include spectral anisotropy, but phase randomness does not allow to include intermittency. Time dependence has been included in a 1D model by Lepreti et al.~\cite{lepreti06}, where time variation is obtained by means of an associated shell model. A minimal Lagrangian map method has been proposed by Rosales \& Meneveau~\cite{rosales06,rosales08} to reproduce 3D hydrodynamic turbulence, and a recent generalization to the MHD case has been presented by Subedi et al.~\cite{subedi14}. Finally, a method to reconstruct a 3D magnetic turbulence with nearly constant magnetic field intensity and a prescribed spectrum has been proposed by Roberts~\cite{roberts12}. 

In this paper we present a new model of synthetic turbulent field, which belongs to the class of ``wavelet-based'' models~\cite{juneja94}. Our model has many aspects similar to the model by Cametti et al.~\cite{cametti98}, but with important differences. The model by Cametti et al. suffers for strong limitations due to large memory requirements when increasing the range of spatial scales. In our model we employ a different algorithm which allows us to reproduce very large ranges of spatial scales with very low memory requirements and short computational times. This feature is very important in all the cases where a high-Reynolds number turbulence is to be represented, as typically happens in astrophysical applications. A more detailed discussion on this point will be given in the next section. Our model generates a solenoidal, time-independent, three-component turbulent vector field within a 3D spatial domain. The field is obtained as a superposition of ``basis functions'' at different spatial scales and positions, whose amplitude is determined through a multiplicative process based on the $p$-model technique~\cite{meneveau87}. It can reproduce both isotropic and anisotropic spectra; in the latter case we can also obtain the kind of anisotropy that can be inferred from the so-called critical balance principle that has been formulated for strong Alfv\'enic turbulence~\cite{goldreich95}. The synthetic field reproduces the high-order statistics of a turbulent field, in particular intermittency. Finally, the field is analytically calculated at any spatial position without employing spatial grids; this feature is particularly useful for test-particle applications because no interpolation processes are required during the calculation of particle evolution. 

The plan of the paper is the following: in Section 2 the synthetic turbulence model is described in details; in Section 3 we validate the model by analyzing its statistical properties; in Section 4 we present the conclusions.


\section{Synthetic turbulence model}
\label{Section:model}
Our synthetic turbulence model generates a three-component solenoidal time-independent turbulent field which will be denoted by ${\bf v}={\bf v}({\bf x})=(v_x,v_y,v_z)$. The field is defined within a 3D spatial domain in form of a parallelepiped $D=\lbrace {\bf x}=(x,y,z)\rbrace=\lbrack 0,L_x\rbrack \times \lbrack 0,L_y\rbrack \times \lbrack 0,L_z\rbrack$. Periodicity is imposed on all the boundaries of the domain $D$. To simulate the turbulent cascade, the field ${\bf v}$ is obtained through a suitable superposition of localized ``basis functions'', each of which represents an eddy characterized by its spatial scale $\ell$, position, amplitude and spatial profile. The scales $\ell$ have discrete values $\ell_m$ which span a range corresponding to the inertial range of the turbulence. The amplitudes of the eddies 
are derived taking into account both their relationship with the average spectral energy flux and the intermittent character of the local energy flux. This is 
simulated through a multiplicative process similar to that used in the $p$-model~\cite{meneveau87}. A detailed description of the model is given in the following.

\subsection{Spectrum and cell hierarchy}
To simulate the process of the eddy breaking within the turbulent cascade, we build a hierarchy of cells at different spatial scales. Each scale is identified by the (integer) index $m=0,\dots,N_s$, where $N_s$ is the number of scales included in the model. Each cell roughly corresponds to the support of a localized function 
representing an eddy (see below). At the largest scale, identified by the index $m=0$, there is only one cell, which coincides with the whole domain $D$; thus, the corresponding typical size is $L_0=(L_x L_y L_z)^{1/3}$. The cells at the next scale $m=1$ are obtained by dividing all the edges of $D$ in two equal parts, thus obtaining eight equal parallelepipeds, each occupying 1/8 of the volume of $D$. Such a process is recursively repeated a number $N_s$ of times. Thus, at the $m$-th scale the cell size is 
\begin{equation}\label{lxlylz}
\ell_{x,m}=2^{-m} L_x \;\;,\;\; \ell_{y,m}=2^{-m} L_y \;\;,\;\; \ell_{z,m}=2^{-m} L_z
\end{equation}
along $x$, $y$ and $z$, respectively, with $m=0,\dots N_s$. At the $m$-th scale, the domain is divided into $2^{3m}$ cells, each occupying a volume $V_m=2^{-3m} L_0^3$ and with a typical size 
\begin{equation}\label{ellm}
\ell_m=(\ell_{x,m}\, \ell_{y,m}\, \ell_{z,m})^{1/3} = 2^{-m} L_0
\end{equation}
Note that every cell at any scale has the same aspect ratio as the domain $D$; this feature will be relaxed in the case of anisotropic spectrum, as explained in Subsection~\ref{subsection:anisotropy}. At any given scale $m$, all the cells form a 3D lattice filling the whole domain $D$. We indicate the cells by
\begin{equation}\label{cell}
C^{(i,j,k;m)}=\lbrace (x,y,z)\rbrace = 
\left[ (i-1) \ell_{x,m}, i\,\ell_{x,m} \right] \times
\left[ (j-1) \ell_{y,m}, j\,\ell_{y,m} \right] \times
\left[ (k-1) \ell_{z,m}, k\,\ell_{z,m} \right] 
\end{equation}
Hereafter the indexes $i,j,k=1,\dots ,2^m$ will identify the cell position within the 3D lattice at the $m$-th scale. The total number of cells contained in the model is indicated by 
\begin{equation}\label{Ncell}
N_{cell}=\sum_{m=0}^{N_s} 2^{3m}
\end{equation}
Note that in the models by Juneja et al.~\cite{juneja94} and Cametti at al.~\cite{cametti98} no cell hierarchy is used because each eddy can occupy any position within the spatial domain.

Cells at the smallest scale have a size of the order of $\ell_{N_s}=2^{-N_s}L_0$. We assume that the eddy amplitudes are non-vanishing in the range of scales $\ell_I \le \ell_m \le \ell_d$, where $\ell_I=2^{-m_I}$ and $\ell_d = \ell_{N_s}$ correspond to the energy injection scale and to the dissipative scale, respectively. Such a range represents the inertial range of the turbulence. In order to have statistical homogeneity, the injection scale $\ell_I$ must be sufficiently smaller than the largest scale $\ell_0$; we set $m_I=2$, corresponding to $\ell_I/\ell_0 = 1/4$. An important parameter of the model is the spectral width $r$ defined as the ratio 
\begin{equation}\label{r}
r=\ell_I/\ell_d=2^{N_s-m_I}
\end{equation}
Within the inertial range, the mean fluctuation amplitude $\Delta v_m$ at the scale $\ell_m$ follows a power law
\begin{equation}\label{deltav}
\Delta v_m = \Delta v_I \left( \frac{\ell_m}{\ell_I}\right)^h
\end{equation}
where $\Delta v_I$ is the fluctuation amplitude at the injection scale $\ell_I$. The exponent $h$ is equal to $1/3$ in the case of a Kolmogorov spectrum. As usual, an expression for the dissipative scale can be found by imposing that at the scale $\ell_d$ the nonlinear time $\tau_{nl}(\ell)=\ell/\Delta v(\ell)$ is equal to the dissipative time $\tau_d(\ell)=\ell^2/\nu$, where $\nu$ is the dissipative coefficient. Using the relation~(\ref{deltav}), this gives
\begin{equation}\label{ld}
\ell_d \sim \left( \frac{\nu }{\Delta v_I}\right)^{\frac{1}{1+h}} \ell_I^{\frac{h}{1+h}} =
\frac{\ell_I}{Re^{\frac{1}{1+h}}}
\end{equation}
where $Re = \Delta v_I \, \ell_I/\nu$ is the Reynolds number. From equation~(\ref{ld}), using the relation~(\ref{r}) the Reynolds number $Re$ can be related to the ratio $r$ and to the parameters of the model:
\begin{equation}\label{Re}
Re \sim \left( \frac{\ell_I}{\ell_d} \right)^{1+h} = r^{1+h} = 2^{(N_s-m_I)(1+h)}
\end{equation}
The tests of the model described in the next section have been performed using $N_s=16$. This corresponds to a spectral width $r=2^{14}\simeq 1.6\times 10^{4}$ giving a spectrum more than 4 decades wide. Using equation~(\ref{Re}) with $h=1/3$ this gives an estimation for the Reynolds number 
$Re \sim 2^{56/3} \simeq 4\times 10^5$. This value of $Re$ is more than two orders of magnitude larger than what can be typically reached in 3D direct simulation with present day standard computational resources.

\subsection{Eddy structure}
The turbulent field is modeled as a superposition of spatially-localized eddies. Each eddy is associated with a cell, so that the total number of eddies coincides with $N_{cell}$. We indicate by $\Delta {\bf v}^{(i,j,k;m)}$ the field of the eddy associated with the cell $C^{(i,j,k;m)}$. Since the field is solenoidal, we write it in terms of a vector potential ${\bf \Psi}^{(i,j,k;m)}$:
\begin{equation}\label{dvPhi}
\Delta {\bf v}^{(i,j,k;m)}({\bf x})=\nabla \times {\bf \Psi}^{(i,j,k;m)}({\bf x}) = 
a^{(i,j,k;m)}\, \nabla \times {\bf \Phi}^{(i,j,k;m)}({\bf x})
\end{equation} 
where the vector function ${\bf \Phi}^{(i,j,k;m)}$ determines the spatial form of the field $\Delta {\bf v}^{(i,j,k;m)}$. We choose the order of magnitude of $\nabla \times {\bf \Phi}^{(i,j,k;m)}$ such as
\begin{equation}\label{orderrot}
|\nabla \times {\bf \Phi}^{(i,j,k;m)}({\bf x})| \sim 1
\end{equation}
for any scale $m$. With this choice, the quantity $a^{(i,j,k;m)}$ in equation~(\ref{dvPhi}) represents the amplitude of the eddy. Both $\Delta {\bf v}^{(i,j,k;m)}({\bf x})$ and ${\bf \Phi}^{(i,j,k;m)}({\bf x})$ are defined in the sub-domain 
\begin{eqnarray}\label{Dsub}
&&D^{(i,j,k;m)}=\lbrace (x,y,z)\rbrace = \nonumber \\
&&\left[ \left(i-\frac{3}{2}\right) \ell_{x,m}, 
\left(i+\frac{1}{2}\right) \ell_{x,m} \right] \times 
\left[ \left(j-\frac{3}{2}\right) \ell_{y,m}, 
\left(j+\frac{1}{2}\right) \ell_{y,m} \right] \times \\
&&\left[ \left(k-\frac{3}{2}\right) \ell_{z,m}, 
\left(k+\frac{1}{2}\right) \ell_{z,m} \right] \nonumber
\end{eqnarray}
and are vanishing outside $D^{(i,j,k;m)}$. Thus, $D^{(i,j,k;m)}$ represents the support of the functions $\Delta {\bf v}^{(i,j,k;m)}$ and ${\bf \Phi}^{(i,j,k;m)}$.
Comparing equations (\ref{cell}) and (\ref{Dsub}) we see that the sub-domain $D^{(i,j,k;m)}$ is wider than the corresponding cell $C^{(i,j,k;m)}$ by a factor 2 along each space direction. Thus, the fields of adjacent cells partially overlap. Indeed, if $D^{(i,j,k;m)}$ and $C^{(i,j,k;m)}$ were coincident, the fluctuating field at a given scale would vanish at any surface border of adjacent cells; this would introduce an artificial periodicity at all the scales that would affect statistical homogeneity. Eddy overlapping is implemented in order to avoid this problem.

\begin{figure}
\begin{center}
\includegraphics[scale=0.75]{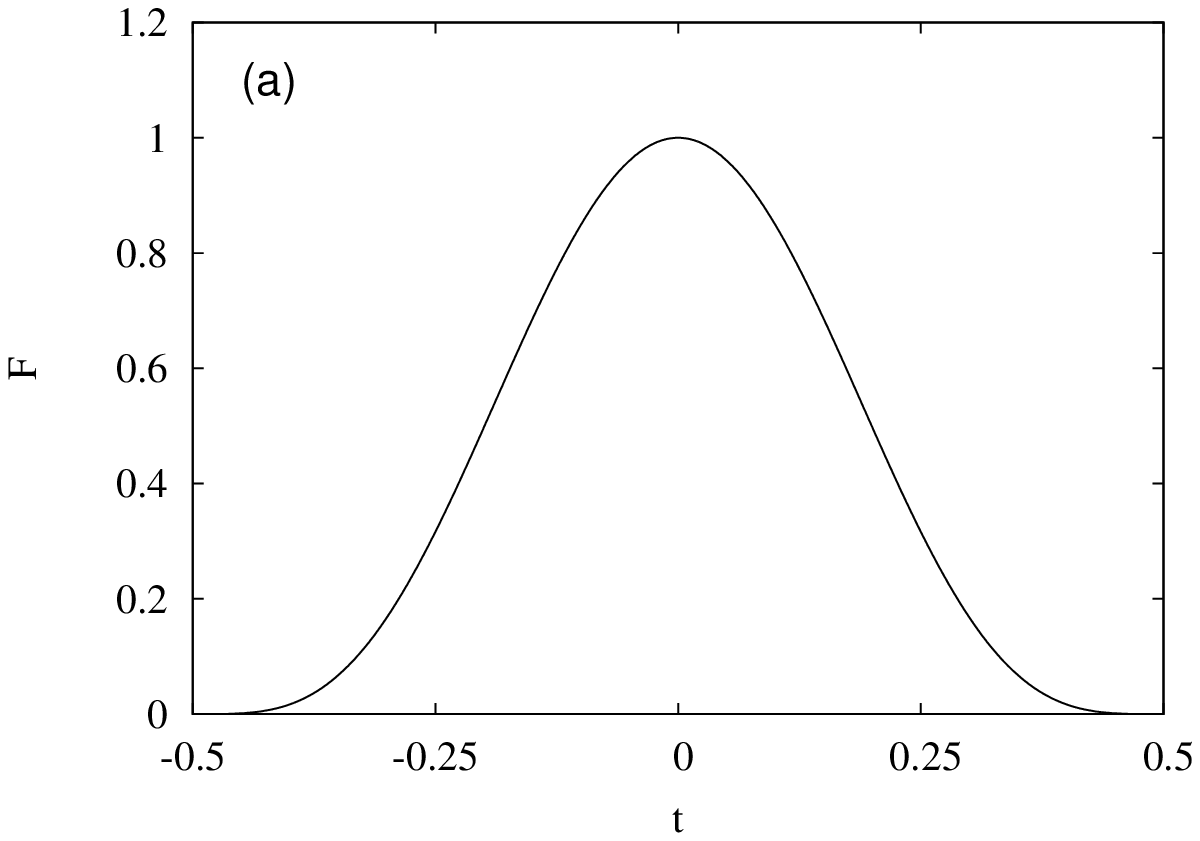}\\
\hskip 12pt \includegraphics[scale=0.72]{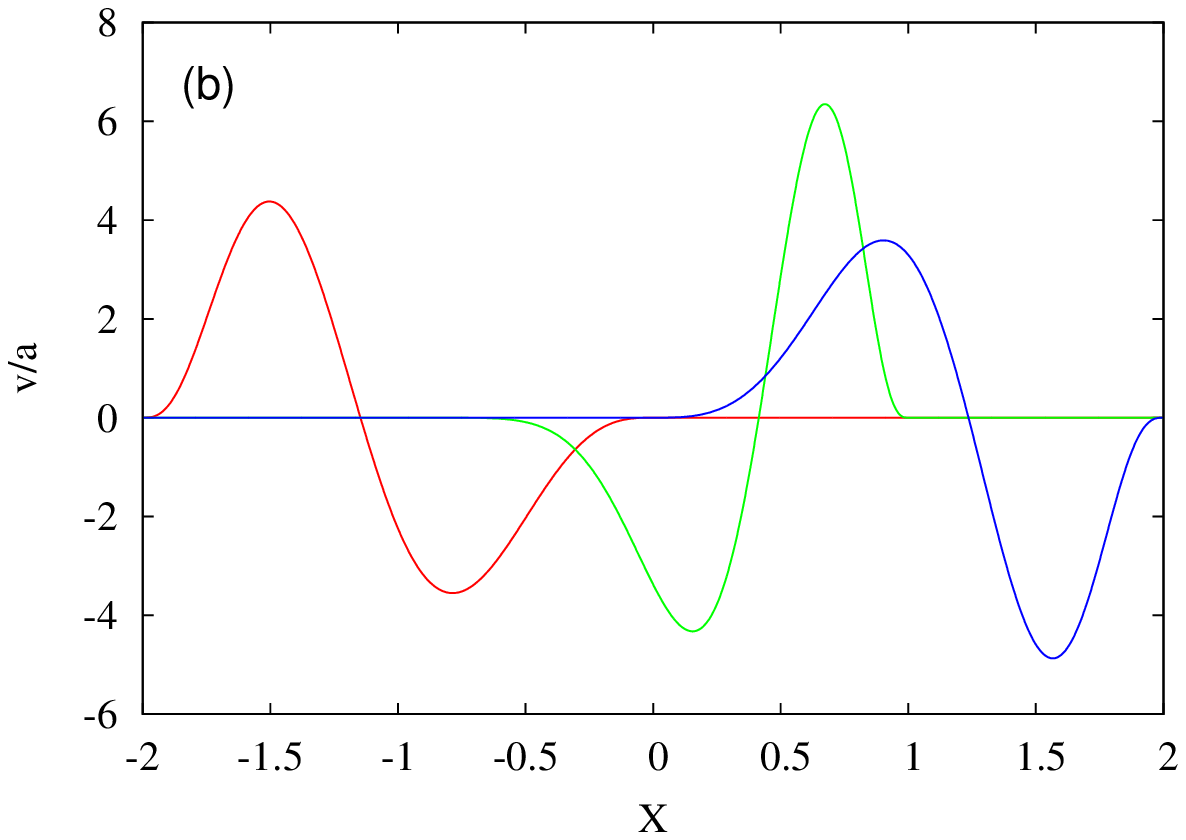}
\caption{Top panel ($a$): the profile of the undistorted base function $F(t)$. Bottom panel ($b$): an example of superposition of neighboring distorted eddies in a given interval along the $x$ direction.}
\label{fig:F}
\end{center}
\end{figure}

Within a given sub-domain $D^{(i,j,k;m)}$ a set of linearly rescaled local spatial coordinates are defined by the relations:
\begin{eqnarray}\label{rescaled}
X^{(i;m)} = X^{(i;m)}(x) = 
\frac{1}{2\,\ell_{x,m}}\left[ x-\left( i-\frac{1}{2}\right)\ell_{x,m}\right] \nonumber \\
Y^{(j;m)} = Y^{(j;m)}(y) = 
\frac{1}{2\,\ell_{y,m}}\left[ y-\left( j-\frac{1}{2}\right)\ell_{y,m}\right] \\
Z^{(k;m)} = Z^{(k;m)}(z) = 
\frac{1}{2\,\ell_{z,m}}\left[ z-\left( k-\frac{1}{2}\right)\ell_{z,m}\right] \nonumber
\end{eqnarray}
The origin $(X^{(i;m)},Y^{(j;m)},Z^{(k;m)})=(0,0,0)$ of rescaled coordinates corresponds to the center of the sub-domain $D^{(i,j,k;m)}$, while each rescaled coordinate varies in the interval $\lbrack -1/2, 1/2 \rbrack$ when the point $(x,y,z)$ varies inside $D^{(i,j,k;m)}$. The explicit form of the vector function ${\bf \Phi}^{(i,j,k;m)}$ is given in terms of the rescaled coordinates by the following expression
\begin{equation}\label{vecpot}
{\bf \Phi}^{(i,j,k;m)}(x,y,z)= \frac{\ell_m}{L_0}\, F(\xi^{(i,j,k;m)})\,F(\eta^{(i,j,k;m)})\,F(\zeta^{(i,j,k;m)})
\end{equation}
where the variables $\xi^{(i,j,k;m)}$, $\eta^{(i,j,k;m)}$ and $\zeta^{(i,j,k;m)}$ are defined below (equations (\ref{xietazeta})), and $F(t)$ is a polynomial function which determines the spatial profile of the eddy. We used the form: 
\begin{eqnarray}\label{Ft}
&&F(t) = 256t^8-256t^6+96t^4-16t^2+1 \;\;, \;\; {\rm for}\;\; -\frac{1}{2} \le t \le \frac{1}{2} \nonumber \\
&&F(t) = 0 \;\;, \; {\rm elsewhere} \nonumber
\end{eqnarray}
A plot of the function $F(t)$ is given in panel ($a$) of Fig. \ref{fig:F}. The function $F(t)$ has one single maximum at $t=0$ ($F(0)=1$) and vanishes with its derivatives up to the $4$-th order at $t=\pm 1/2$. Then, equation~(\ref{vecpot}) corresponds a localized eddy which matches with neighboring eddies with continuous derivatives up to the $4$-th order. This implies that the turbulent field is continuous with all its derivatives up to the third order; in particular, the vorticity (if we interpret ${\bf v}$ as a velocity field) or the current density (if we interpret ${\bf v}$ as a magnetic field) are continuous with their first derivatives. 
This feature is different from what done in the models by Juneja et al.~\cite{juneja94} and Cametti at al.~\cite{cametti98}, in which the profile of the eddy is simpler (a piecewise-linear function), but discontinuities are present in the first derivatives of the turbulent field. The choice of having a more regular field has mainly been done in the perspective of employing the model in test-particle studies; this is useful, for instance, if a term proportional to the current density (the resistive electric field) is included in the motion equation of particles. We also note that $F(t) \sim 1$, in the interval $-1/2 \le t \le 1/2$.

The variables $\xi$, $\eta$ and $\zeta$ are related to the rescaled coordinates by the nonlinear relations:
\begin{eqnarray}\label{xietazeta}
\xi^{(i,j,k;m)} = X^{(i;m)}+\gamma_x^{(i,j,k;m)} \left( {X^{(i;m)}}^2-\frac{1}{4} \right) \nonumber \\
\eta^{(i,j,k;m)} = Y^{(j;m)}+\gamma_y^{(i,j,k;m)} \left( {Y^{(j;m)}}^2-\frac{1}{4} \right) \\
\zeta^{(i,j,k;m)} = Z^{(k;m)}+\gamma_z^{(i,j,k;m)} \left( {Z^{(k;m)}}^2-\frac{1}{4} \right) \nonumber
\end{eqnarray}
where $\gamma_x^{(i,j,k;m)}$, $\gamma_y^{(i,j,k;m)}$, $\gamma_z^{(i,j,k;m)}$ are constants which are randomly chosen in the interval $\lbrack -1,1\rbrack$. The nonlinear mapping (\ref{xietazeta}) introduces a distortion in the spatial profile of the eddy along the three spatial directions, whose entity is determined by the three random numbers $\gamma_x^{(i,j,k;m)}$, $\gamma_y^{(i,j,k;m)}$ and $\gamma_z^{(i,j,k;m)}$. This effect has been introduced in order to improve the statistical homogeneity of the fluctuating field. Note that the above regularity properties of the vector potential are preserved by the mapping (\ref{xietazeta}). A plot illustrative of the profile of few distorted and overlapped eddies is given in panel ($b$) of Fig. \ref{fig:F}. Finally, using the definitions (\ref{lxlylz}), (\ref{ellm}), (\ref{rescaled}), and (\ref{xietazeta}), it can be verified that the form (\ref{vecpot}) of the vector function ${\bf \Phi}^{(i,j,k;m)}$ satisfies the assumption (\ref{orderrot}).

\subsection{Turbulent cascade and intermittency}
The amplitudes $a^{(i,j,k;m)}$ of the eddies are determined considering the phenomenology of the turbulent cascade. In a stationary situation, the mean energy transfer rate $\langle \epsilon \rangle$ at a given spatial scale $\ell$ is independent of $\ell$~\cite{K41}, angular parentheses indicating a spatial average. For hydrodynamic turbulence $\langle \epsilon \rangle \sim \lbrack \Delta v (\ell) \rbrack^3/\ell$, implying that the mean fluctuation at the scale $\ell$ is $\Delta v (\ell) \propto \ell^{1/3}$. This scaling law corresponds to the Kolmogorov spectrum, where the spectral energy density is $e(k) \propto k^{-5/3}$, with $k$ the wavenumber. 
In general, we assume that in the inertial range $\Delta v$ follows the power law given in equation~(\ref{deltav}), corresponding to $e(k) \propto k^{-(2h +1)}$.
However, it turns out from experimental observation that the energy transfer rate $\epsilon$ is not spatially uniform, but rather change from place to place according to the effectiveness of nonlinear couplings~\cite{k62}. Consequently, the amplitude of fluctuations is not spatially uniform, but fluctuations stronger than the average value $\langle \Delta v (\ell) \rangle$ form locally, which are separated by regions of weaker fluctuations. This feature propagates to smaller scales through a multiplicative process, becoming more and more relevant with decreasing $\ell$. Thus, at small scales the field is characterized by very strong and localized fluctuations with wide ``quiet'' regions in between: this is the phenomenology of intermittency. 

\begin{figure}
\begin{center}
\includegraphics[bb=20 85 700 370,scale=0.6]{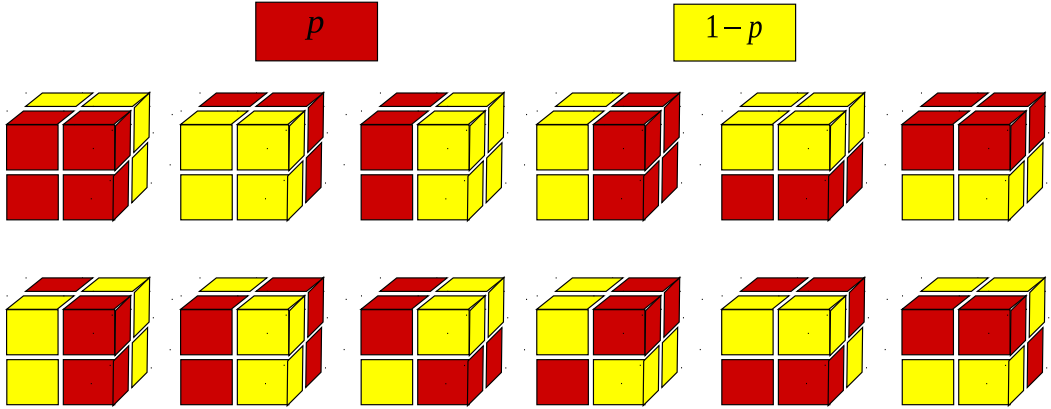}\\
\caption{A graphic representation of the twelve "heritage patterns". In each pattern, cells which receive 
more (less) energy are represented in red (yellow).}
\label{fig:sketch}
\end{center}
\end{figure}

In our model such a process is modeled as in the $p$-model by Meneveau \& Sreenivasan~\cite{meneveau87}, where $p$ is a fixed parameter chosen in the interval $\lbrack 1/2,1\rbrack$. Energy flows from large to smaller eddies with an unequal rate $\epsilon$: each ``parent'' eddy at a scale $\ell_m$ gives energy to its eight ``daughter" eddies at the scale $\ell_{m+1}$ with two possible rates; namely, $\epsilon_{m+1}=2p\epsilon_m\ge \epsilon_m$ for four daughter eddies and $\epsilon_{m+1}=2(1-p)\epsilon_m\le \epsilon_m$ for the remaining four daughter eddies. For $p=1/2$ we have $\epsilon_{m+1}=\epsilon_m$, i.e. the rate $\epsilon$ is equal at all the scales and positions; this corresponds to a non-intermittent fluctuating field. With increasing $p$ above the value $1/2$, differences between the rates increase and the level of intermittency increases, as well. In our synthetic turbulence model $p$ is a free parameter that we use to investigate the effects of intermittency. 
More specifically, the transfer rate is recursively determined for the eddies daughters of the ``$(i,j,k;m)$'' parent eddy by: 
\begin{equation}\label{epsilon}
\epsilon_{m+1,n}=2p\,\epsilon_m\,\beta_n^{(i,j,k;m)}+2(1-p)\,\epsilon_m\,(1-\beta_n^{(i,j,k;m)}), 
\;\;\; m=0,\dots,N_s\;\;, \;\; n=1,\dots,8
\end{equation}
where $\beta_n^{(i,j,k;m)}=1$ for four randomly chosen daughters (for instance, $n=3,5,7,8$) who receive more energy, while $\beta^{(i,j,k;m)}=0$ for the remaining four daughters ($n=1,2,4,6$) who receive less energy.
The choice of the four daughter eddies which will receive more energy and of those which will receive less 
energy is made among twelve possible "heritage patterns", which are schetched in Fig.~\ref{fig:sketch}. 

Finally, the amplitude of any eddy is given by 
\begin{equation}\label{amp}
a^{(i,j,k;m)}=\sigma^{(i,j,k;m)} a_0 \left[\frac{\epsilon_m^{(i,j,k;m)}}{\epsilon_0}
\frac{\ell_m}{\ell_0}\right]^{h}
\end{equation}
where $a_0=a^{(1,1,1;0)}$ and $\epsilon_0=\epsilon^{(1,1,1;0)}$ are the amplitude and the energy transfer rate at the largest scale, respectively, and the exponent $h$ is related to the spectral slope. 
The quantity $\sigma^{(i,j,k;m)}$ in equation~(\ref{amp}) represents the sign of the eddy and it is randomly chosen as $\sigma^{(i,j,k;m)}=1$ or $\sigma^{(i,j,k;m)}=-1$.
In conclusion, the turbulent field is given by
\begin{equation}\label{flucfield}
{\bf v} ({\bf x})= \sum_{m=m_I}^{N_s}\, \sum_{i,j,k=1}^{2^m} 
\nabla \times {\bf \Psi}^{(i,j,k;m)} ({\bf x}) =
\sum_{m=m_I}^{N_s}\, \sum_{i,j,k=1}^{2^m} 
a^{(i,j,k;m)} \, \nabla \times {\bf \Phi}^{(i,j,k;m)} ({\bf x})
\end{equation}
where the derivatives in the $\nabla$ operator are to be calculated with respect to the coordinates $x$, $y$ and $z$, and the index $m_I$ identifies the injection scale $\ell_I$. 
Using the expressions given in equations (\ref{rescaled})-(\ref{xietazeta}), the analytical form of all the quantities appearing in the equation~(\ref{flucfield}) can be explicitly calculated.

\subsection{Eddy superposition algorithm}
Equation~(\ref{flucfield}) gives the turbulent field as a superposition of fluctuating fields, each one associated with a particular eddy. As mentioned before, in our model the total number of eddies coincides with the number $N_{cell}$ of cells, given in equation~(\ref{Ncell}). Then, the number of eddies exponentially increases with the number of scales $N_s$ included in the model; for instance, using $N_s=16$ we have $N_{cell} \simeq 3\times 10^{14}$, which is a very large number of eddies. The storage of the whole information defining all the eddies in the computer memory for high values of $N_{cell}$ would represent a difficulty because of large memory requirements. This is the case for example in the 3D model by Cametti et al.~\cite{cametti98}, where the position of each eddy is randomly chosen within the spatial domain; as a result, the memory requirement exponentially increases with $N_s$ and obliges one to use relatively small values for $N_s$, i.e., relatively small spectral widths $r$. In fact, the spectral width considered in the paper~\cite{cametti98} is of the order of two decades. In the present model we use a different algorithm which avoids to use large memory storage power, even for very large values of $N_{cell}$. This allows us to reach larger spectral widths $r$ with a modest computational effort. This aspect is important for having a low-cost synthetic turbulence model, as desirable. In the following we describe how our algorithm is built. 

\noindent
a) In our model eddies are not randomly translated. Thus, the location of the support $D^{(i,j,k;m)}$ of any eddy is known {\em a priori} (equation~(\ref{Dsub})). As a consequence, when calculating ${\bf v}$ at a given spatial point ${\bf x}$, only a small number of terms give a non-vanishing contribution to the sum of equation~(\ref{flucfield}): namely, those terms corresponding to eddies whose support contains the point ${\bf x}$. Taking into account the partial overlapping of neighboring eddies, it can be verified that, for a given position ${\bf x}$ and for a given value of the scale index $m$, only 8 eddies satisfy the following condition
\begin{equation}\label{inclusion}
{\bf x}\in D^{(i,j,k;m)}
\end{equation} 
and then contribute to build the field ${\bf v}$ at the position ${\bf x}$. The algorithm first selects these eddies on the base of the position ${\bf x}$, taking into account the partial overlapping of neighbouring eddies, as well as periodicity in the case eddies are close to the boundaries of the spatial domain. 
We indicate the 8 selected eddies satisfying the condition~(\ref{inclusion}) and belonging to the $m$-th scale by the indexes $^{(\mu;{\bf x};m)}$, with $\mu=1,\dots,8$. Thus, the equation~(\ref{flucfield}) is replaced by
\begin{equation}\label{shortsum}
{\bf v} ({\bf x})= \sum_{m=m_I}^{N_s}\, \sum_{\mu =1}^{8} 
\nabla \times {\bf \Psi}^{(\mu;{\bf x};m)} ({\bf x})
\end{equation}
where ${\bf \Psi}^{(\mu;{\bf x};m)}$ is the vector potential associated to the eddy whose support is $D^{(\mu;{\bf x};m)}$, satisfying the condition~(\ref{inclusion}).
Equation~(\ref{shortsum}) indicates that the number of terms that have to be calculated when evaluating the field at a position ${\bf x}$ is now $N_{term}=8\, N_s$, which is much smaller than $N_{cell}$. Moreover, while $N_{cell}$ increases exponentially with the number $N_s$ of scales, $N_{term}$ is simply proportional to $N_s$. This fact allows for an extremely fast evaluation of the turbulent field, even for large spectral width. For instance, using a number $N_s=16$ of scales, corresponding to a spectral width larger than 4 decades ($r\simeq 1.6 \times 10^4$ with $m_I=2$), only $128$ terms are included in the sum~(\ref{shortsum}). Moreover, increasing the number of scales $N_s$ by a factor 2 would increase the spectral range by a factor $2^{N_s}$ while the computation time would simply be increased by a factor 2.

\noindent
b) As explained above, the vector potential ${\bf \Psi}^{(i,j,k;m)}$ associated with each eddy is characterized by a set of random parameters, that are: (i) $\gamma_n^{(i,j,k;m)}$, defining the distortion of each eddy (equation~(\ref{xietazeta})) ; (ii) the sign $\sigma^{(i,j,k;m)}$ (equation~(\ref{amp})); and (iii) $\beta_l^{(i,j,k;m)}$ defining the energy transfer rate of each eddy in terms of the rate of its parent eddy (equation~(\ref{epsilon})), which, in turn, determines the eddy amplitude. In order to calculate the sum in equation~(\ref{shortsum}) we have to know all these parameters for the $N_{term}$ eddies involved in the sum. 
In principle, this could be done by calculating a-priori these random quantities for all the eddies and storing this information in the computer memory. Then, when a given eddy is involved in the field evaluation, the corresponding quantities could be recalled and used to calculate the field. However, the total number of eddies $N_{cell}$ can be very large; for instance, using a number of scales $N_s=16$ we have $N_{cell} \simeq 3\times 10^{14}$ (equation (\ref{Ncell})). Then, storing the information defining all the eddies would require a huge memory. For that reason, we used a different procedure, which is described in the following.
Since the eddies involved in the sum of equation (\ref{shortsum}) have been selected only on the base of their location with respect to the position ${\bf x}$ (condition (\ref{inclusion})), their defining parameters must depend only on the location of the eddies within the lattice of cells. Such parameters are determined in the following way: for any given cell an integer $\lambda^{(i,j,k;m)}$ is calculated using the expression:
\begin{equation}\label{lambda}
\lambda^{(i,j,k;m)} = i+(j-1)2^m + (k-1)2^{2m}+\nu_m
\end{equation}
where the integer $\nu_m$ is defined as follows:
\begin{equation}
\nu_m=\left\{
\begin{array}{cc}
  0 & \quad \mbox{if} \quad m=0; \\
  \displaystyle {\sum_{n=0}^{m-1} 2^{3m}} & \quad \mbox{if} \quad m \ge 1.
\end{array}\right.
\end{equation}
It can be verified that, for $m$ varying between $0$ and $N_s$ and for $i$, $j$ and $k$ varying between $1$ and $2^m$, the expression (\ref{lambda}) generates all the integers between $1$ and $N_{cell}$. This defines a one-to-one correspondence between the set $\lbrace 1 \le \lambda \le N_{cell}, \; \lambda \; {\rm integer}\rbrace$ and the set of cells. In other words, $\lambda^{(i,j,k;m)}$ represent an absolute address for any cell. 
The integer $\lambda^{(i,j,k;m)}$ is used as a seed for a random number generating routine (RNGR), which 
is called a fixed number $i_{sample}$ of times, with $i_{sample}$ an integer. Finally, the resulting number 
calculated by the RNGR is used to generate the parameters $\gamma_n^{(i,j,k;m)}$, $\sigma^{(i,j,k;m)}$, and 
$\beta_l^{(i,j,k;m)}$ which define the eddy associated to the given cell. In this way, the properties of the 
$N_{term}$ eddies appearing in the sum (\ref{shortsum}) are univocally determined as functions of the given 
position ${\bf x}$. This completely defines all the quantities in equation~(\ref{shortsum}) and allows for an 
explicit evaluation of the field ${\bf v}$ at any spatial position ${\bf x}$. Moreover, different choices of 
the integer $i_{sample}$ gives origin to different realizations of the turbulent field. This allows to build 
an ensemble of configurations for the turbulent field.

Strictly speaking, the parameters $\gamma_n^{(i,j,k;m)}$, $\sigma^{(i,j,k;m)}$ and $\beta_l^{(i,j,k;m)}$ are not random quantities because they are univocally determined as soon as the position ${\bf x}$ has been chosen. On the other hand, the set of possible values of the seed $\lambda^{(i,j,k;m)}$ is formed by $N_{cell}$ of values, which is an extremely large value (equation~(\ref{Ncell})). This fact, in practice, ensures a global randomness of the parameters which define the structure of single eddies. 
We note that in the above-described algorithm nothing needs to be kept in memory: each time the field ${\bf v}$ is to be calculated at a position ${\bf x}$, this is done deducing all the properties of the $N_{term}$ involved eddies directly from their absolute address $\lambda^{(i,j,k;m)}$. 

Finally, it should be pointed out that, at variance with other methods, no spatial grid is used; on the contrary the field is directly calculated at the given spatial point without any interpolation procedure.

\subsection{Anisotropic spectrum}
\label{subsection:anisotropy}
In many examples of real-world flows, the turbulence spectrum is not isotropic in the wave-vector space. For instance, this happens in MHD when a large-scale magnetic field ${\bf B}_0$ is present. In this case, ${\bf B}_0$ introduces a preferential direction and the energy cascade tends to preferentially develop in the directions perpendicular to ${\bf B}_0$. This generates anisotropic spectra both for the velocity and for the magnetic field perturbations, in which perpendicular wave-vectors prevail over parallel ones. This has been shown in theoretical studies (e.g., \cite{shebalin83,carbone90,OughtonEA94}). Moreover, observations indicate that in the solar wind turbulence spectrum the distribution of wave-vectors of magnetic fluctuations has a significant population quasi-perpendicular to the mean magnetic field \cite{matthaeus86,matthaeus90}.


Within that context, Goldreich \& Sridhar~\cite{goldreich95} introduced the principle of ``critical balance''. In that formulation it is assumed that the nonlinear time for an eddy with sizes $\ell_\parallel$ and $\ell_\perp$ (parallel and perpendicular to ${\bf B}_0$, respectively) depends only on the transverse size $\ell_{\perp}$: $\tau_{nl}=\ell_\perp /\Delta a(\ell_\perp)$, $\Delta a(\ell_\perp) \propto \ell_\perp^{1/3}$ being the velocity/magnetic field fluctuation amplitude which is assumed to follow the Kolmogorov scaling law. Moreover, all along the spectrum a balance is assumed to hold between $\tau_{nl}$ and the propagation time $t_A=\ell_\parallel /c_A$, which is the time a perturbation takes to travel over a distance $\ell_\parallel$ along ${\bf B}_0$ at the Alfv\'en velocity $c_A$. 
This gives a relationship between parallel and perpendicular lengths of eddies:
\begin{equation}\label{lperppar}
\ell_\parallel \propto \ell_\perp^{2/3}
\end{equation}
equation~(\ref{lperppar}) indicates that, when going from large to small scales, $\ell_\parallel$ decreases slower than $\ell_\perp$, i.e., structures more and more elongated in the ${\bf B}_0$ direction are found at small scales. This corresponds to a spectrum which is more anisotropic at small scales than at large scales.

We explored the possibility to reproduce the anisotropy corresponding to the critical balance principle by our synthetic turbulence model. This has been done by modifying the above-described cell hierarchy in the following way. First, $z$ has been conventionally chosen as the direction parallel to the background magnetic field ${\bf B}_0$. Second, we introduce the possibility to have anisotropic cell divisions; this means that, when going from the $m$-th scale to the $(m+1)$-th scale, all the cells at the $m$-th scale are divided only along the $x$ and $y$ directions, while no division is performed in the $z$ direction. In other words, the aspect ratio of cells at the $m$-th scale is different from that of cells at the $(m+1)$-th scale, the latter being more elongated along $z$ than the former. In contrast, in the previously-described isotropic cell division, when going from the $m$-th to the $(m+1)$-th scale, the cells are equally divided along all the three spatial directions, keeping the same aspect ratio at all the scales. These two possibilities are described by the equations:
\begin{equation}\label{isoanidiv}
\ell_{x,m+1}=\ell_{x,m}/2 \;\; , \;\; \ell_{y,m+1}=\ell_{y,m}/2 \;\; , \;\;
\ell_{z,m+1}=\ell_{z,m}/\rho_m 
\end{equation}
where $\rho_m=2$ in the case of isotropic division, while $\rho_m=1$ in the case of anisotropic division. The relation~(\ref{lperppar}) between parallel and perpendicular lengths can be reproduced by a suitable choice of the coefficients $\rho_m$ in equation~(\ref{isoanidiv}), given by the following sequence:
\begin{equation}\label{etam}
\lbrace \rho_m,\, m=0,\dots,N_s \rbrace = \lbrace 2,2,1,2,2,1,2,2,1,\dots \rbrace
\end{equation}
corresponding to one anisotropic division every three divisions.

In the anisotropic version of the model, some definitions used in the previously-described isotropic case must be modified accordingly. The $m$-th scale in the $z$ direction (equation~(\ref{lxlylz})) is now defined as $\ell_{z,m}=L_z/\pi_m$, where 
\begin{equation}\label{pim}
\pi_m = \prod_{i=0}^m \rho_i
\end{equation}
The index $k$, which identifies the cell position in the $z$ direction within the lattice (see, e.g., equation~(\ref{cell})), now varies in the interval $k=1,\dots,\pi_m$. Since the smallest size of an eddy is now $\ell_{\perp,m}=\ell_{x,m}=\ell_{y,m}\le \ell_{z,m}$ we now adopt the following expression for the vector function ${\bf \Phi}^{(i,j,k;m)}$ (compare with equation~(\ref{vecpot})):
\begin{equation}\label{vecpotani}
{\bf \Phi}^{(i,j,k;m)}(x,y,z)= \frac{\ell_{\perp,m}}{L_0}\, F(\xi^{(i,j,k;m)})\,F(\eta^{(i,j,k;m)})\,F(\zeta^{(i,j,k;m)})
\end{equation}
Finally, since in the critical balance principle the spectrum is assumed to follow a Kolmogorov law with respect to $k_\perp$, the equation~(\ref{amp}) defining the eddy amplitude is now replaced by 
\begin{equation}\label{ampani}
a^{(i,j,k;m)}=\sigma^{(i,j,k;m)} a_0 \left[\frac{\epsilon_m^{(i,j,k;m)}}{\epsilon_0}
\frac{\ell_{\perp,m}}{\ell_0}\right]^{1/3}
\end{equation}
All the other features of the model and of the algorithm remain unchanged.

It is worth mentioning that the model can be adapted to reproduce other anisotropy types, such as the ones generated in shear, rotating, or wall-bounded flows. Such flexibility makes the model suitable to describe diverse physical systems.


\section{Testing the model}
In order to test the model described in previous Section, the standard diagnostics for the description of intermittent turbulence have been routinely performed on the synthetic data. In this Section we present the results of the analysis.
A number of realizations of the synthetic turbulent field ${\bf v(x)}$ have been generated both for the isotropic and for the anisotropic version of the model. For each run, one single sample was generated with $i_{sample}=1$. The typical Kolmogorov scaling exponent $h=1/3$ was imposed for all runs, while the strength of the intermittency was changed by allowing the parameter $p$ to take the following values: $p=0.5$, corresponding to non-intermittent turbulence; $p=0.7$, a realistic value close to the typical observations in ordinary fluid turbulence; $p=0.9$, representing a ``super-intermittent'' case, which will be mostly used as benchmark for the parametric description of the model. 
The relevant scales were imposed as described in Section~\ref{Section:model} (the integral scale $\ell_I=L_0/4$) or estimated by looking at the spectra (the dissipation scale $\ell_d \simeq L_0/(2\times10^{4})$, see Figure~\ref{fig:spectra}), resulting in the effective Reynolds number 
$Re \sim (\ell_I/\ell_d)^{4/3} \simeq 8.5\times 10^4$, which is smaller but close to the estimation 
given in the equation (\ref{Re}). 
In order to ensure ergodicity, ten independent synthetic trajectories of length $L=40\ell_I$ were extracted from each run as one-dimensional samples, with spatial resolution $dr\simeq 1.5\times 10^{-5}\ell_I$ chosen as to ensure the inclusion of the whole inertial range in the spectrum. For each trajectory, the longitudinal field increments $\Delta v$ were computed at different scales $l$, $\langle v\rangle$ and $\sigma_v$ being respectively their mean and standard deviation. Since from now on we will only consider the component of the field along the virtual trajectory, we will simplify the notation by defining $v(s)\equiv {\bf v(x)}\cdot {\bf \hat{s}}$, where ${\bf \hat{s}}$ is the unit versor of the trajectory.
Successively, the following quantities have been obtained for each run: (1) the autocorrelation function $A_c(l)=\langle [v(s)-\langle v\rangle][v(s+l)-\langle v\rangle)]\rangle/\sigma_v^2$, which gives useful information about the correlation scale of the field, $l_c$; (2) the associated energy power spectrum $E(k)$ ($k=2\pi/l$ being the wave-vector associated with the scale $l$), whose power-law scaling exponent has to be compared with the one imposed for the model field fluctuations, $h$; (3) the Probability Distribution Functions (PDFs) of the scale-dependent increments, $P(\Delta v)$, whose deviation from Gaussian will qualitatively illustrate the presence of intermittency; (4) the structure functions $S_q(l)=\langle |\Delta v|^q\rangle \sim l^{\zeta_q}$, i.e. the scale-dependent $q$-th order moments of the field increment distribution, and their anomalous scaling exponents $\zeta_q$; (5) the kurtosis $K=S_4/S_2^2$, an alternative, quantitative measure of intermittency (fully determined by the scaling of the structure functions), along with its scaling exponent $\kappa$; (6) and, finally, a box-counting based multifractal analysis, providing some finer detail on the geometrical properties of the flow.

It should be noticed that the present version of our model does not include the skewness of the PDFs, a crucial ingredient of intermittency universally observed in real turbulence~\cite{FRISCH}. For this reason, it will be necessary to use the absolute value of the fluctuations to prevent the odd-order structure functions to vanish.

\subsection{Isotropic turbulence}
\begin{figure}
\begin{center}
\includegraphics[scale=0.4]{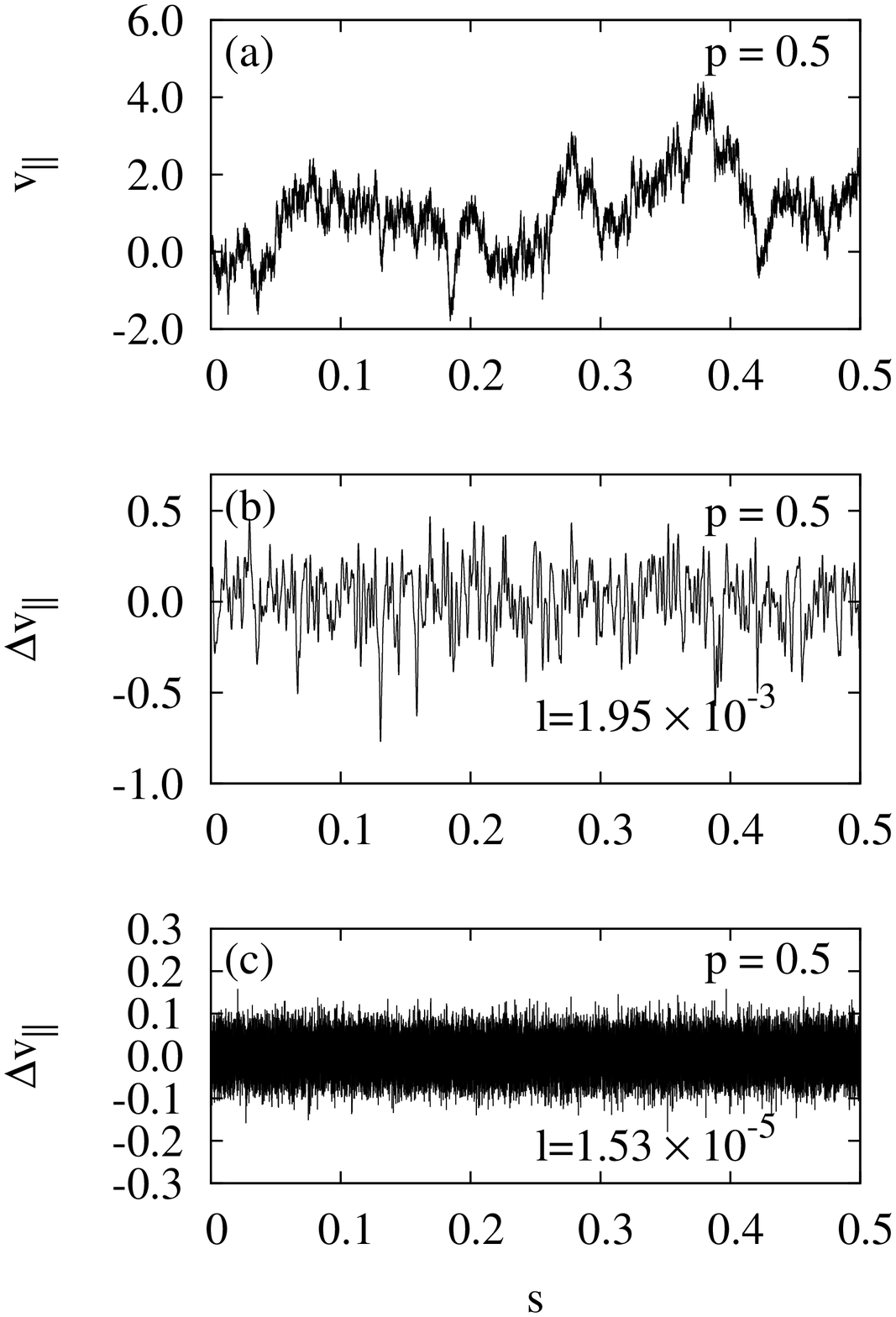}
\includegraphics[scale=0.4]{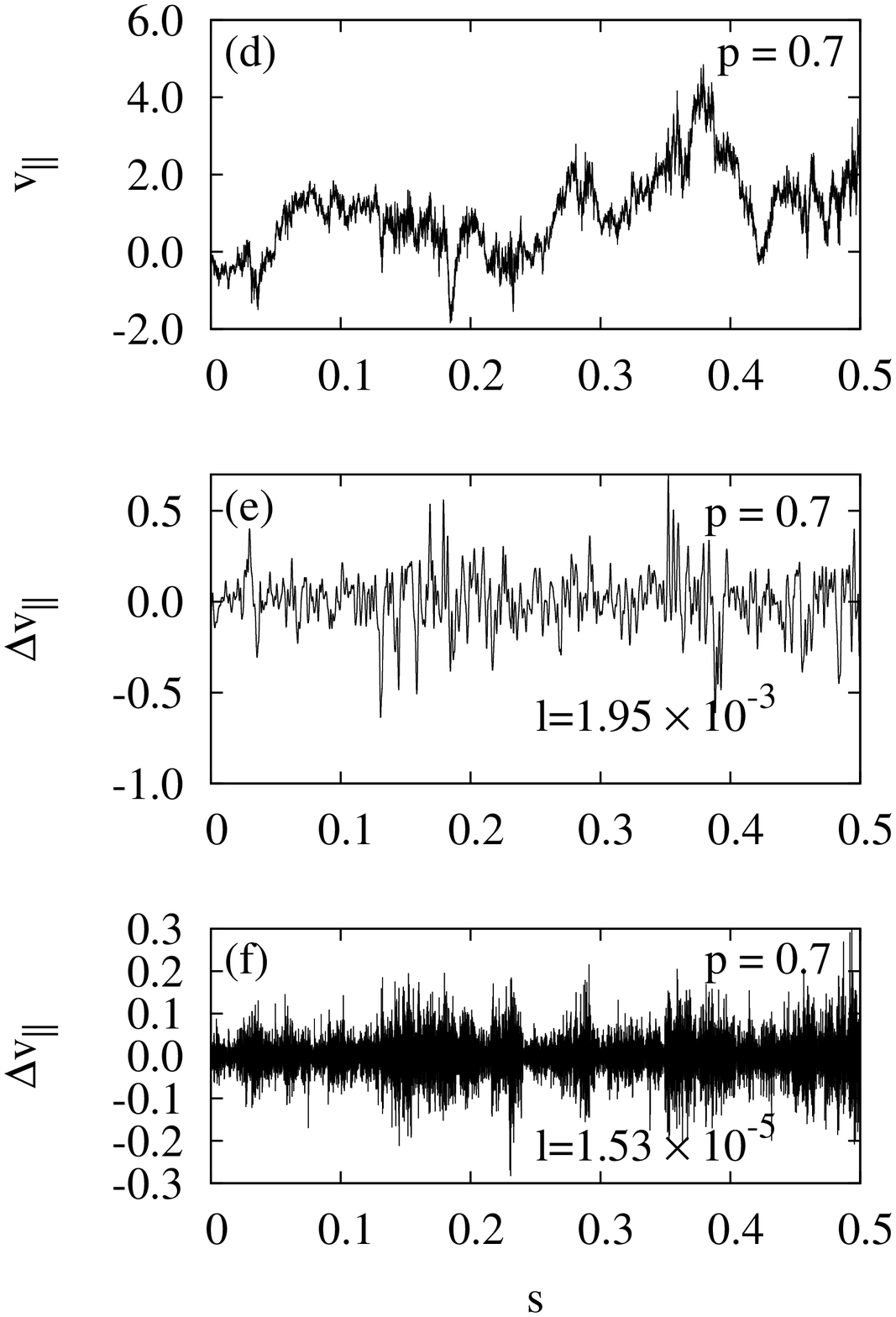}
\caption{Examples of profile of the longitudinal component of the field $v$, together with the increments $\Delta v$ evaluated at two different scales (see legend). Left panels ($a$, $b$, $c$): no intermittency ($p=0.5$); right panels ($d$, $e$, $f$): standard intermittency ($p=0.7$).}
\label{fig:data}
\end{center}
\end{figure}

Examples of the field longitudinal component $v(s)$, extracted from one of the realizations of isotropic turbulence, is shown in the top panels of Figure~\ref{fig:data} for two values of the intermittency parameter $p$. Along with the longitudinal field component, the increments $\Delta v$ at two different scales $l$ are included in the figure. The presence of intermittency is revealed by the scale-dependent general properties of the increments, and in particular by their increasing burstiness towards smaller scales.
%
\subsubsection{Two-dimensional Spectrum.}
\begin{figure}
\begin{center}
\includegraphics[scale=1.]{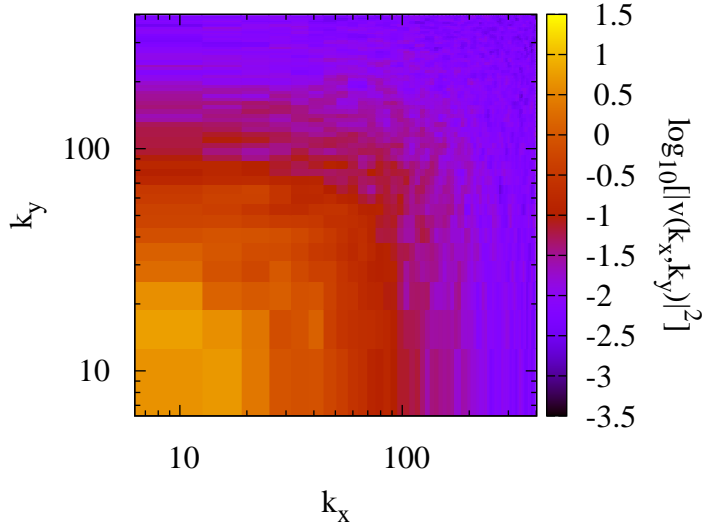}
\caption{Isocontours of the two-dimensional spectrum in the plane $k_x,k_y$ for the total power associated to the intermittent field ${\bf v}$. The image refers to the case $p=0.7$. Similar results hold for the other levels of intermittency (not shown).}
\label{fig:spectrum2d}
\end{center}
\end{figure}

%
%
For the isotropic runs, a preliminary study of the full spectral properties of the fields revealed the presence of a weak residual anisotropy, probably due to the shape of the generating functions. Indeed, the two-dimensional cut of the spectrum presented in Figure~\ref{fig:spectrum2d} displays an excess of power along the diagonals, which results in roughly squared rather than circular isocontours. This feature is consistently observed in all of the three two-dimensional spectral cuts (not shown). In order to mitigate this weak deviation from isotropy, and to increase the statistical significance of the sample, for each realization ten different trajectories were selected at varying angles with the domain axes, so that the solid angle was homogeneously sampled. Each sample was analyzed separately using the tools described above. The results were finally averaged over the ten different samples from all the trajectories. The corresponding standard deviation was used as an estimate of the uncertainty in the model parameters. 
%
%
\subsubsection{Autocorrelation Function.}
\begin{figure}
\begin{center}
\includegraphics[scale=0.5]{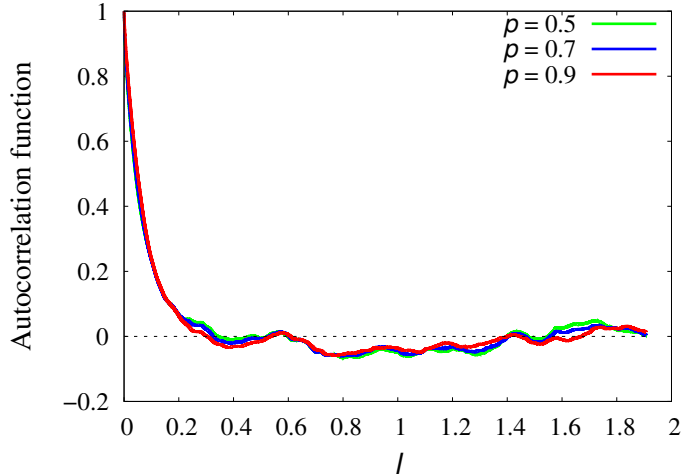}
\caption{The autocorrelation function for the longitudinal field component, for the three values of the intermittency parameter $p$.}
\label{fig:autocorrelation}
\end{center}
\end{figure}

%
%
Figure~\ref{fig:autocorrelation} shows examples of the autocorrelation function versus the separation scale $l$, for different values of the intermittency parameter $p$. The autocorrelation functions display the typical behavior for turbulent fields, with a parabolic decay near the origin (not shown). A faster, quasi-exponential decay follows toward large separation, where eventually the small-amplitude fluctuations around zero determine the noise level.
As customary, an estimate of the correlation scale can be obtained as the scale at which the autocorrelation function reaches the uncorrelated-scale noise level. The values obtained for the three cases are collected in Table~\ref{Table:p}, and are consistent with the imposed integral scale $\ell_I=L_0/4$. There is no relevant difference between the three runs, as intermittent corrections to the autocorrelation function are expected to be small. 
%
%
\subsubsection{Omnidirectional spectrum.}
\begin{figure}
\begin{center}
\includegraphics[scale=0.5]{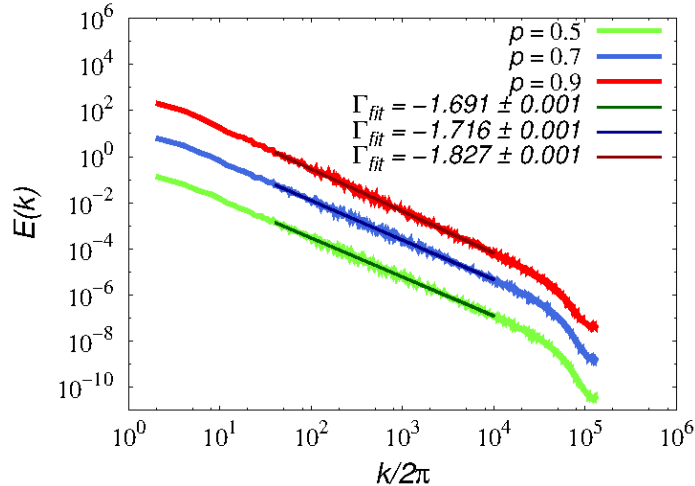}
\caption{The one-dimensional power spectra $E(k)$ for the longitudinal component of the synthetic field, for the three runs. Power-law fits are also superposed. The scaling exponents are collected in Table~\ref{Table:p}, showing good agreement with the imposed Kolmogorov-like spectrum.}
\label{fig:spectra}
\end{center}
\end{figure}

%
%
For all runs, the energy power spectra evaluated along each trajectory and then averaged, provide quick information about the scaling properties of the fluctuations, and are given in Figure~\ref{fig:spectra}, along with power-law fits in the inertial range. At small scale, a quasi-exponential decay indicates the smoothness of the field, due to the differentiability of the mother functions, and mimicking the dissipation scale of turbulence. On the contrary, at very large scales the absence of correlation weakly flattens the spectrum. The spectral indexes obtained from the power-law fit within the inertial range are listed in Table~\ref{Table:p}. For all runs, the exponents are slightly larger than the values expected using the simple relation $\Gamma=2h+1$, with the input parameter $h=1/3$. This is evident for the case $p=0.5$, for which $\Gamma=1.69$ instead of $5/3$. Such weak discrepancy is consistently observed for the other two runs with $p\neq0.5$, when considering the intermittent correction.
%
%
\subsubsection{Probability Distribution Functions of longitudinal increments.}
\begin{figure}
\begin{center}
\includegraphics[scale=0.5]{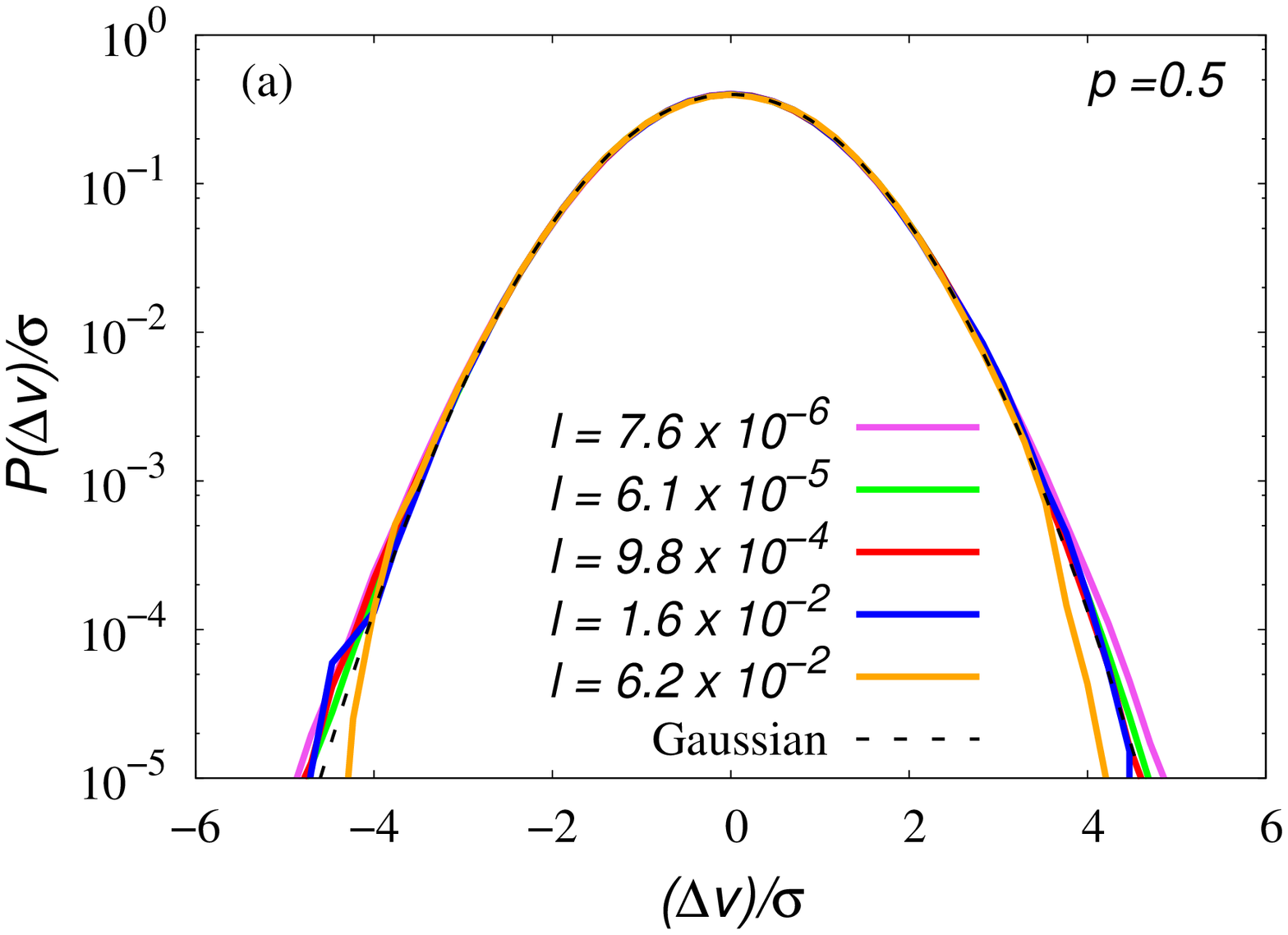}
\includegraphics[scale=0.5]{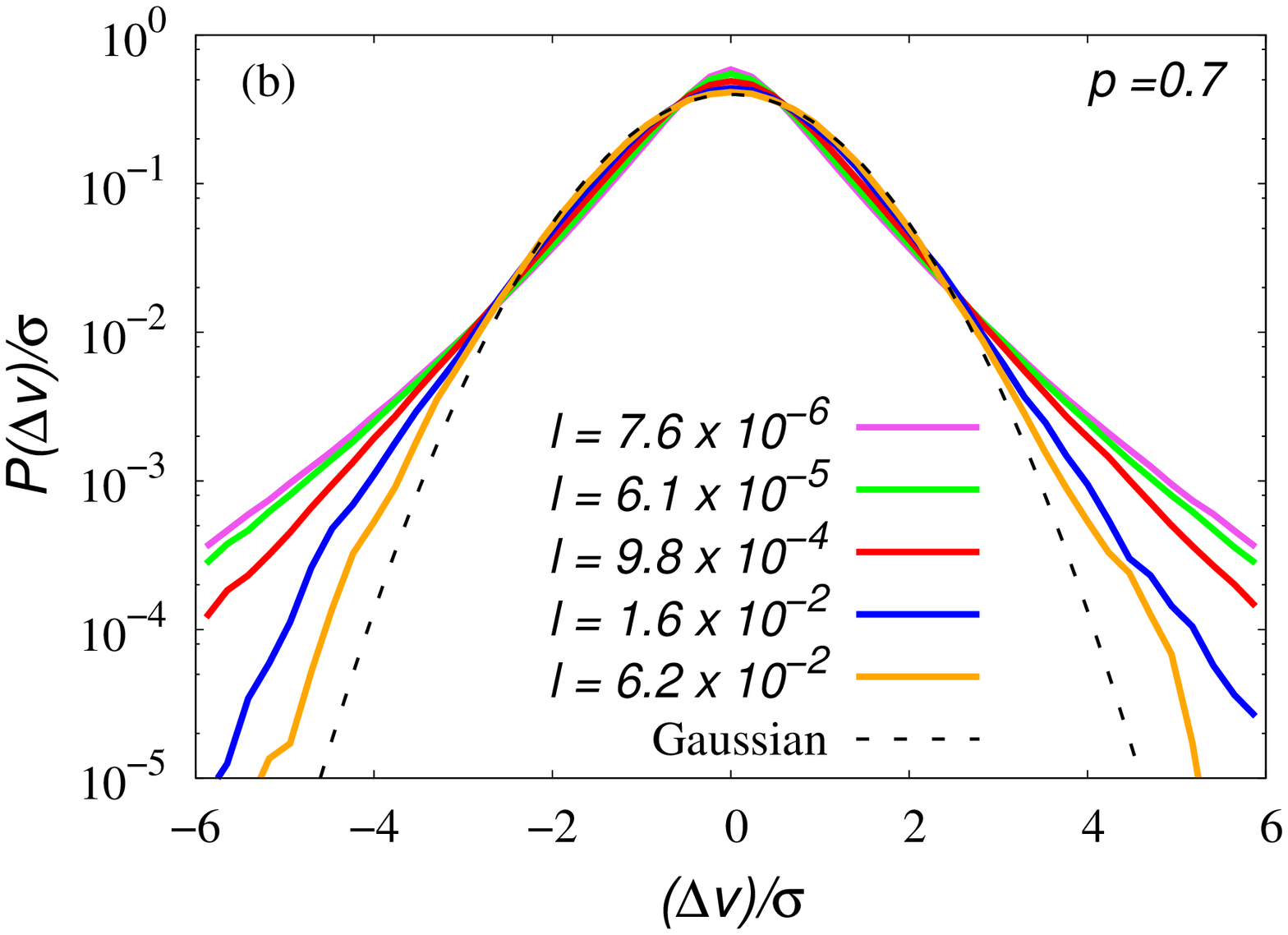}
\includegraphics[scale=0.5]{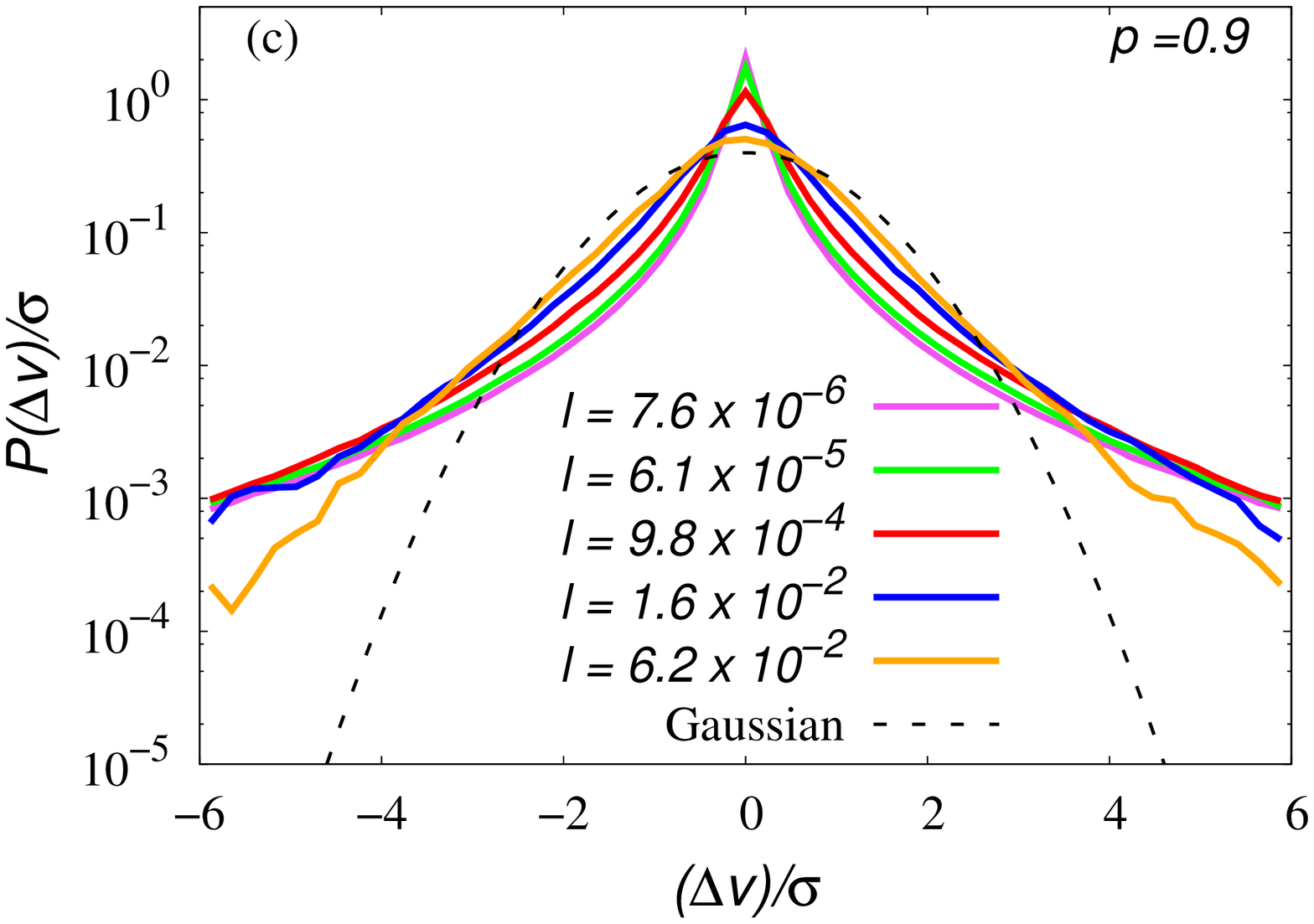}
\caption{Probability Distribution Functions of the standardized field increments on different scales (see legend) for the three runs. Top panel ($a$): $p=0.5$; central panel ($b$): $p=0.7$; bottom panel ($c$): $p=0.5$.}
\label{fig:pdf}
\end{center}
\end{figure}

%
%
In order to account for inhomogeneities of the energy flux in the cascade process, i.e. of intermittency, examples of the increment PDFs at different scales are collected in Figure~\ref{fig:pdf} for three values of $p$. The increments have been previously standardized for each scale, in order to allow a proper comparison. It is evident that in the absence of intermittency ($p=0.5$) the distribution functions are roughly Gaussian, and almost identical at all scales. This indicate self-similarity of the fluctuations and is the result of an homogeneous redistribution of the energy along the cascade. For ``realistic'' values of the intermittent parameter ($p=0.7$), the typical increase of the distribution tails toward small scales is observed~\citep{FRISCH}. This captures the increasing localization of energy as the scale decrease, spontaneously arising in turbulent flows and well reproduced by the model. The ``super-intermittent'' case ($p=0.9$) shows even more evident high tails of the distributions (not shown in this paper).
%
%
\subsubsection{Structure Functions.}
\begin{figure}
\begin{center}
\includegraphics[scale=0.5]{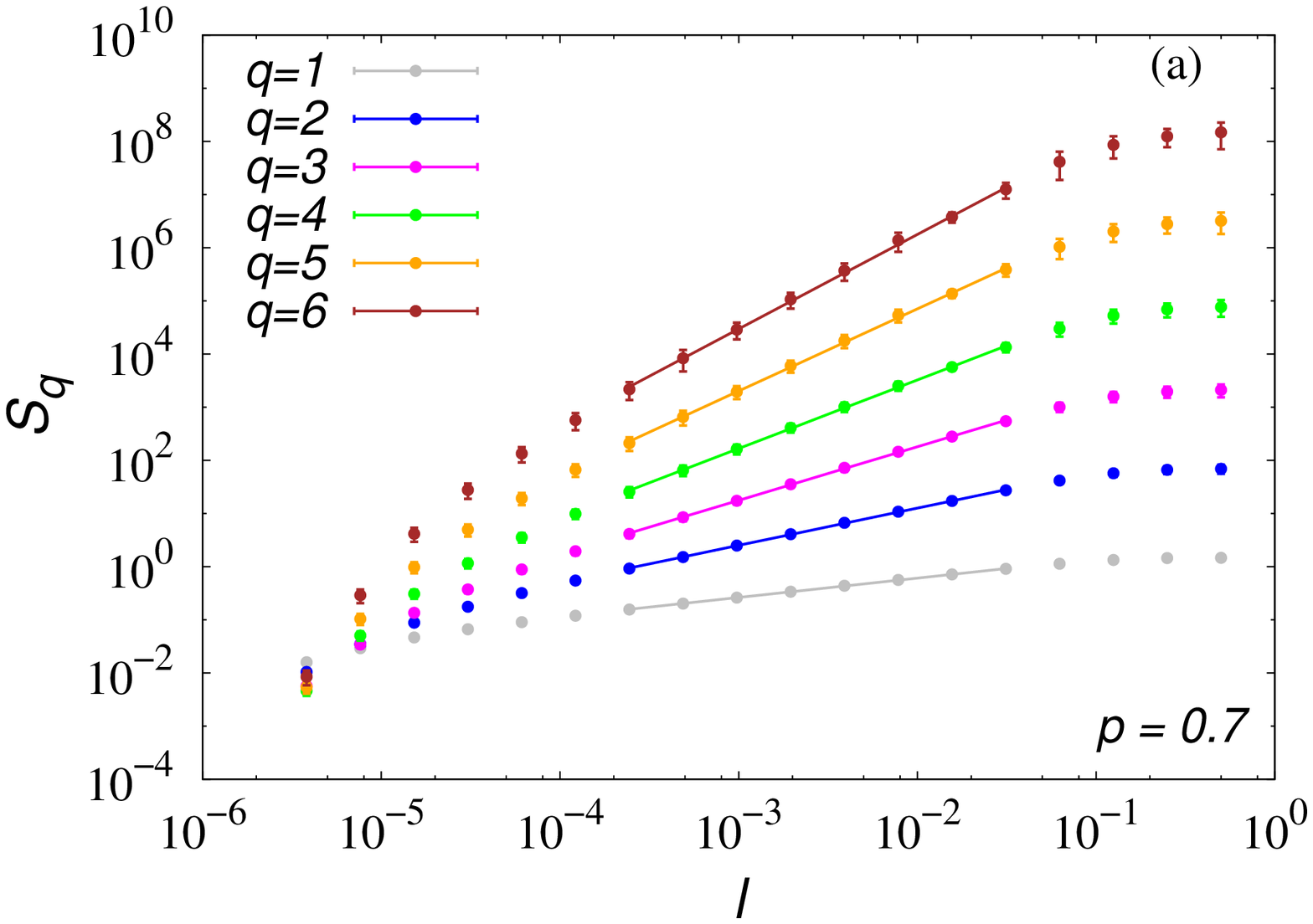}
\includegraphics[scale=0.5]{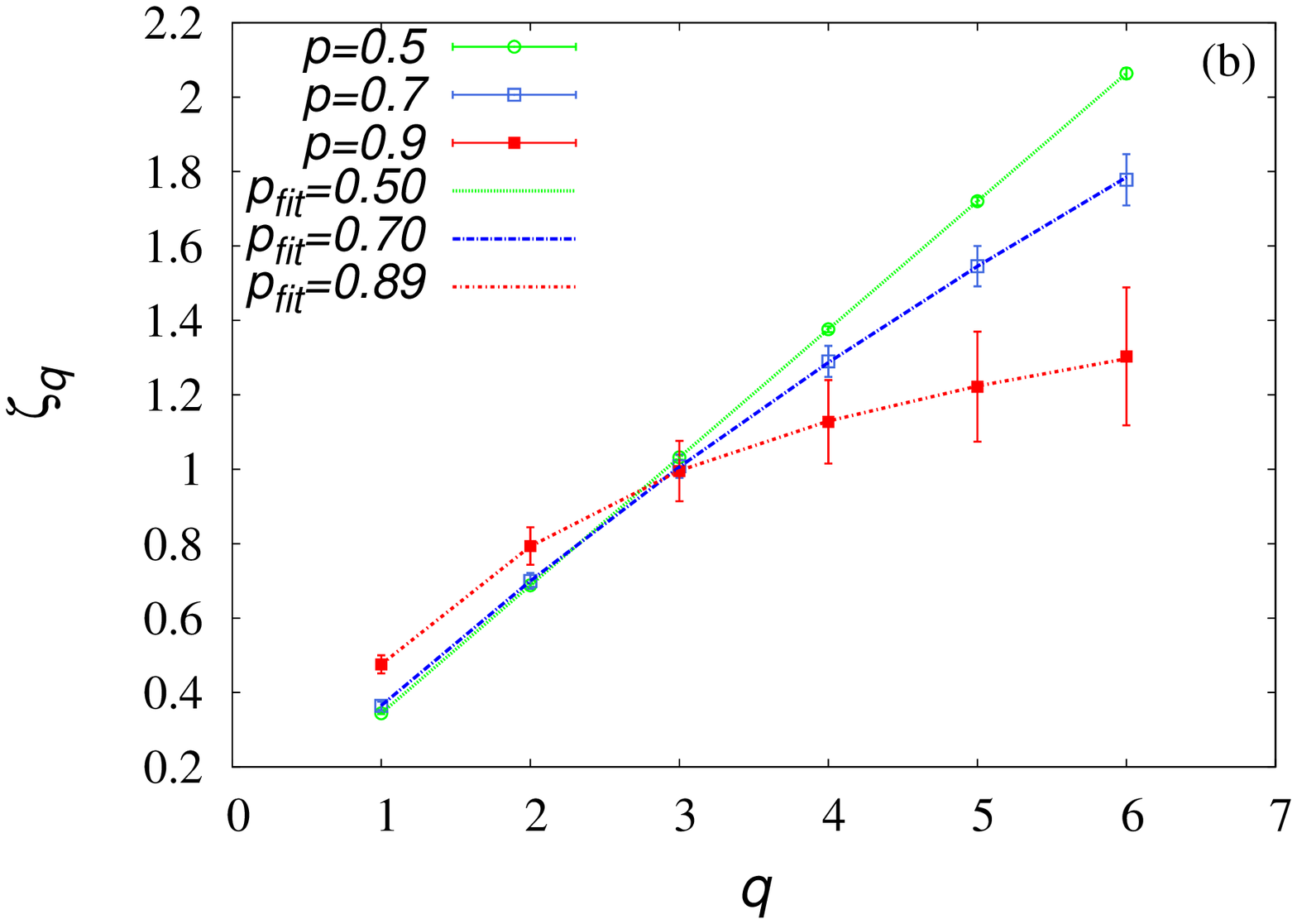}
\caption{Top panel ($a$): the structure functions for the longitudinal component of the field for the case $p=0.7$. Power-law fits used to evaluate the scaling exponents $\zeta_q$ are superimposed. Bottom panel ($b$): The anomalous scaling of the structure functions, highlighted by the nonlinear order dependency of the scaling exponents $\zeta_q$, for three values of $p$. Fits with the $p$-model, equation~(\ref{pmodel}), are indicated as lines. The agreement of the data with the model is excellent.}
\label{fig:structurefunctions}
\end{center}
\end{figure}

%
%
An alternative description of the intermittency is obtained by means of the anomalous scaling of the structure functions $S_q(l)$. Examples are shown in panel ($a$) of Figure~\ref{fig:structurefunctions} for the realistic intermittency case $p=0.7$, for orders up to $q=6$ (convergence of the moments has been tested following~\citet{ddewit,ddewit2}). In the intermediate range of scales, roughly corresponding to the spectral inertial range, the structure functions have been fitted to power laws. The resulting scaling exponents are collected in panel ($b$) of Figure~\ref{fig:structurefunctions} for the three different values of the parameter $p$. Their deviation from the linear prediction $\zeta_q \sim h q$ identifies the effects of intermittency. For a more quantitative estimate, the scaling exponents have been fitted to a $p$-model~\cite{meneveau87}, whose prescription gives 
\begin{equation}
\zeta_q =1-\log_2\left[ p^{h q}+(1-p)^{h q}\right] \, .
\label{pmodel}
\end{equation}
The fitting curves are indicated in the figure as lines, showing good agreement with the data. The corresponding empirical intermittency parameters $p_{fit}$ are collected in Table~\ref{Table:p}, and are consistent with the prescribed values. This confirms that the model is able to effectively generate the desired degree of intermittency in the data by adjusting the parameter $p$. 
%
%
\subsubsection{Kurtosis.}
\begin{figure}
\begin{center}
\includegraphics[scale=0.5]{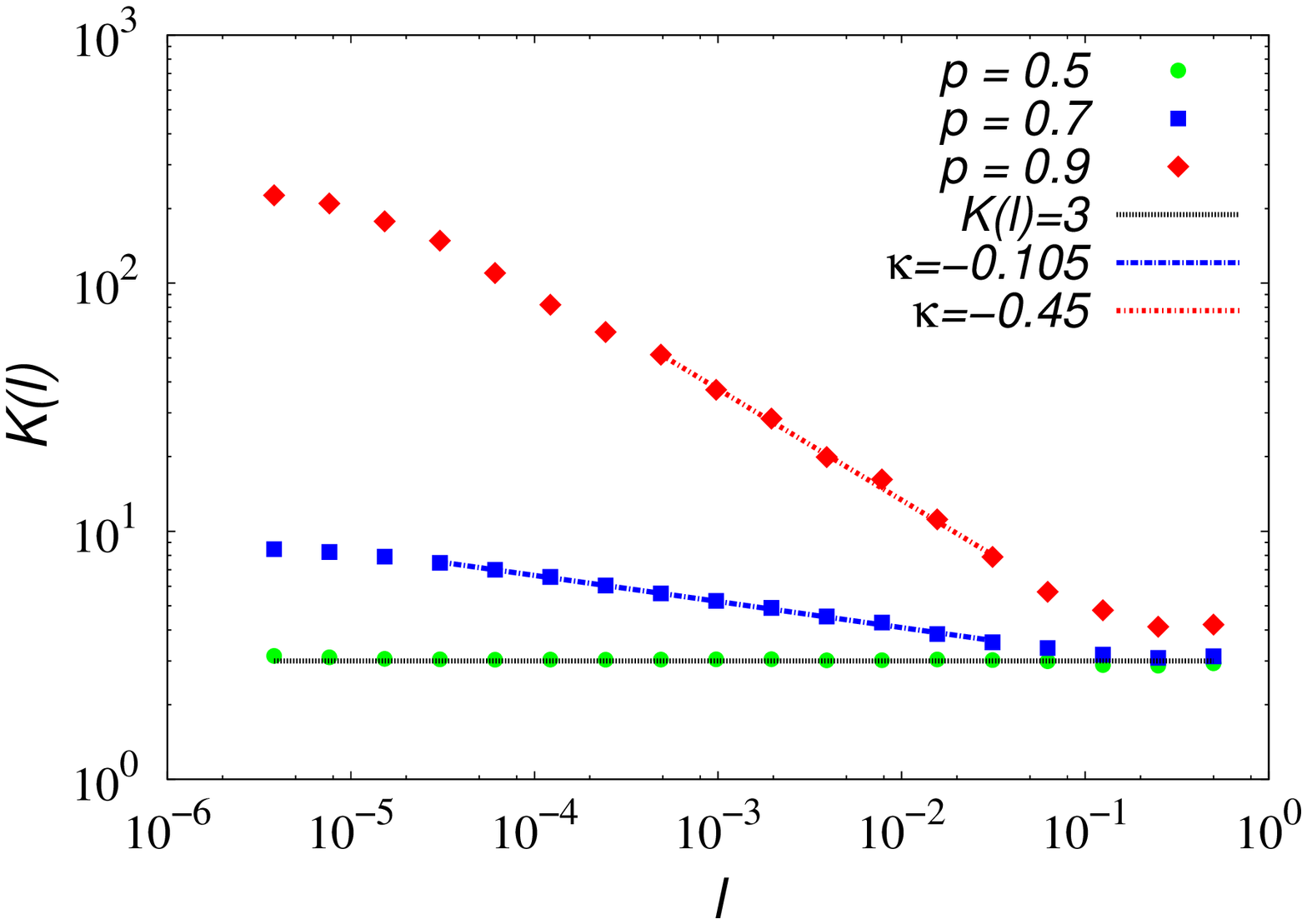}
\caption{The scaling dependence of the kurtosis $K$, for the three values of the intermittency parameter. The Gaussian value $K=3$ is indicated, as well as a power-law fit in the inertial range for the two intermittent cases.}
\label{fig:kurtosis}
\end{center}
\end{figure}

%
%
Figure~\ref{fig:kurtosis} shows the scaling behavior of the kurtosis $K(l)$ for the three values of $p$. The non-intermittent run gives the constant value $K=3$ at all scales, as expected for a Gaussian variable. When intermittency is included, the kurtosis is Gaussian at large scales, roughly down to the correlation scale, and increases toward small scales as a power law $K(l)\sim l^{-\kappa}$. In Navier-Stokes turbulence, it is often observed that $\kappa\simeq0.1$ (also described by the $p$-model and by the She-L\'ev\^eque model), which is consistent with the value obtained by fitting the case $p=0.7$. As expected, saturation of the kurtosis is evident for scales smaller than the dissipative scale $\ell_d$. Furthermore, note that the largest kurtosis attained by the model in the realistic intermittency case ($k_{max} \simeq 10$) is compatible with the values normally found in many experimental observations with a comparable inertial range extension (or Reynolds number). For the case with $p=0.9$, the scaling exponent of the kurtosis is larger, consistent with a more efficient intermittency.
%
%
All the diagnostics described above shows that the synthetic data are consistent with the prescribed values of power spectral decay and intermittency. This demonstrates that the data are representative of a real-world, tunable turbulence, and can therefore be used for numerical studies.
\begin{table}
\begin{center}
\caption{For the three isotropic runs with different intermittency levels $p$, we show here: the correlation length $l_c$, as estimated from the autocorrelation function; the spectral index $\Gamma$, as obtained fitting the spectrum with a power-law; the empirical value of the parameter $p_{fit}$, as obtained from a $p$-model fit of the structure functions scaling exponents; and the scaling exponent of the kurtosis $\kappa$, as estimated through a power-law fit. For the case $p=0.5$, the value $\kappa=0$ was assumed without fitting the kurtosis.}    
\vskip 12pt
\begin{tabular}{ccccc} 
\hline
\hline 
     $p$     &        $l_c$        &       $\Gamma$         &       $p_{fit}$     &          $\kappa$       \\ 
\hline
\hline
 ~~ $0.5$ ~~ &  ~~$0.21\pm0.07$ ~~ & ~~ $1.691 \pm0.005$  ~~ & ~~ $0.5 \pm0.1 $ ~~ & ~~      $ 0 $     ~~ \\ 
 ~~ $0.7$ ~~ &  ~~$0.21\pm0.06$ ~~ & ~~ $1.716 \pm0.001$  ~~ & ~~ $0.71\pm0.02$ ~~ & ~~ $0.101\pm0.006$ ~~ \\ 
 ~~ $0.9$ ~~ &  ~~$0.20\pm0.04$ ~~ & ~~ $1.827 \pm0.001$  ~~ & ~~ $0.89\pm0.02$ ~~ & ~~ $0.42 \pm0.03$  ~~ \\ 
\hline
\hline
  \end{tabular}
  \label{Table:p}
  \end{center}
  \end{table}
%
\subsubsection{Multifractal Analysis.}
\begin{figure}
\begin{center}
\begin{minipage}{175mm}
{\resizebox*{8.3cm}{!}{\includegraphics{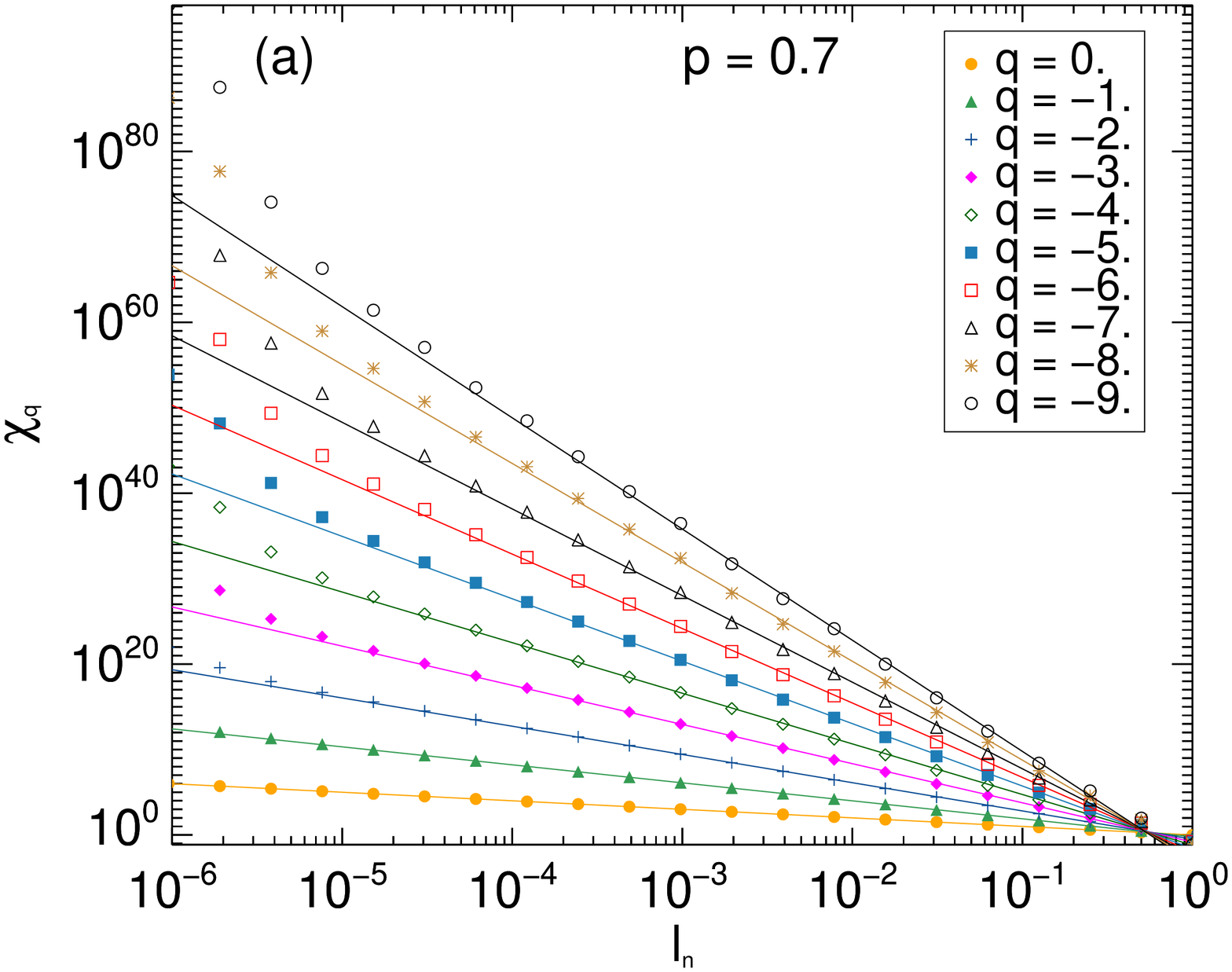}}}%
{\resizebox*{8.3cm}{!}{\includegraphics{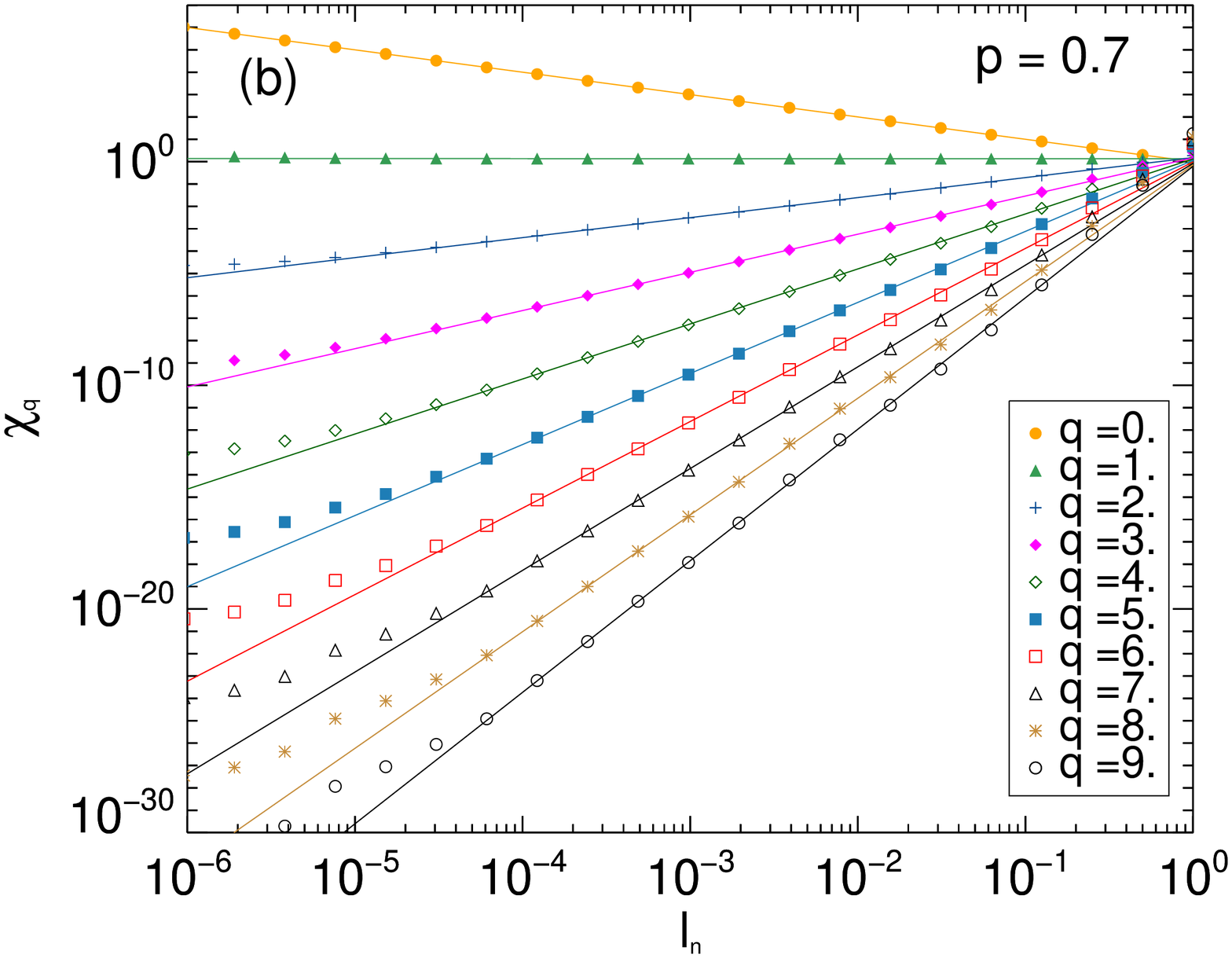}}}%
\caption{
The partition functions $\chi_q$ ($ q < 0 $ left panel ($a$), $ q > 0 $ right panel ($b$)) for the case $p=0.7$. Power law fits (solid lines) are performed in a wide range of scales, roughly corresponding to the spectral inertial range.}
\label{fig:funz_partiz} 
\end{minipage}
\end{center}
\end{figure}

\begin{figure}
\begin{center}
\includegraphics[width=8cm]{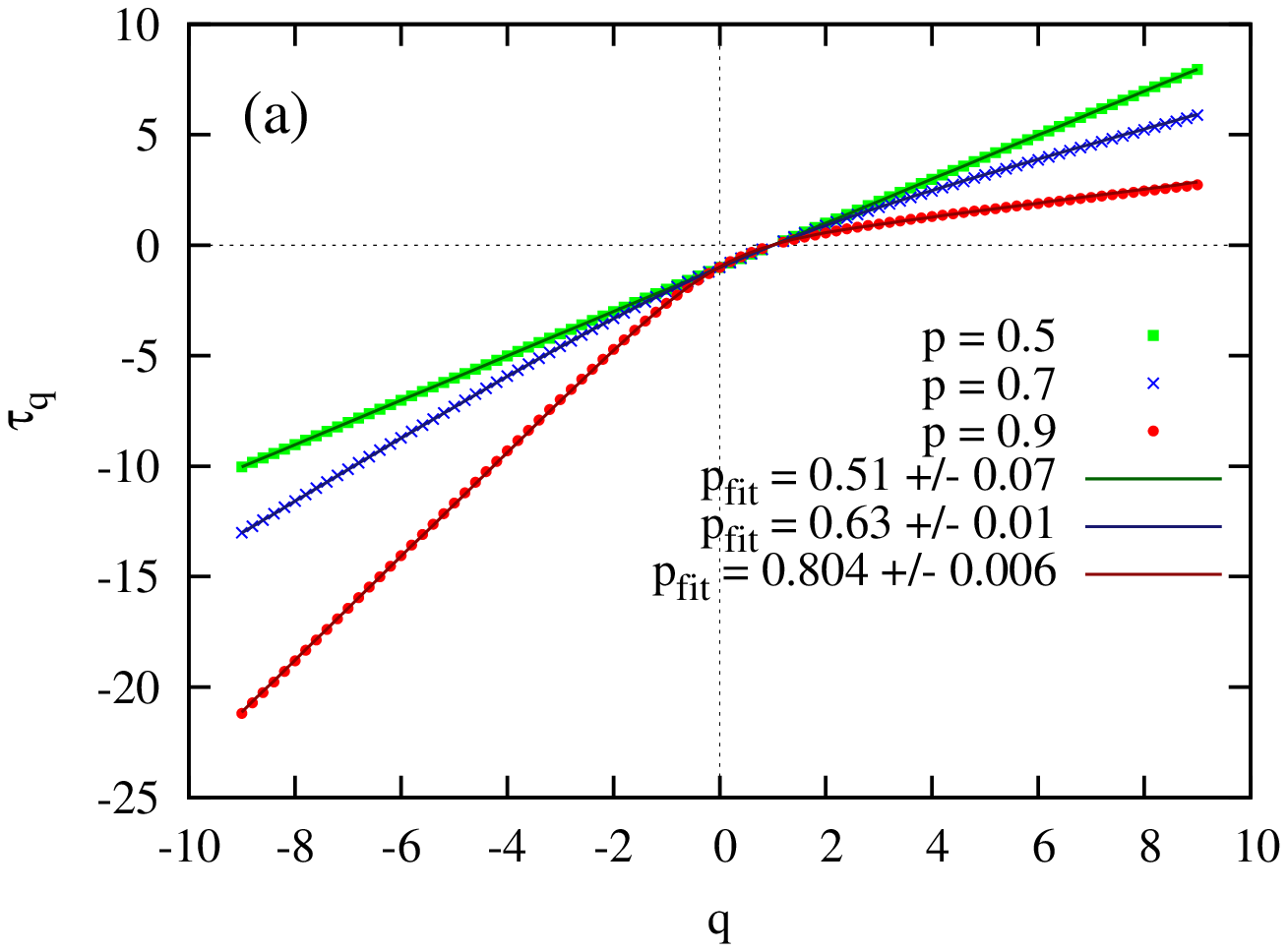}
\includegraphics[width=8cm]{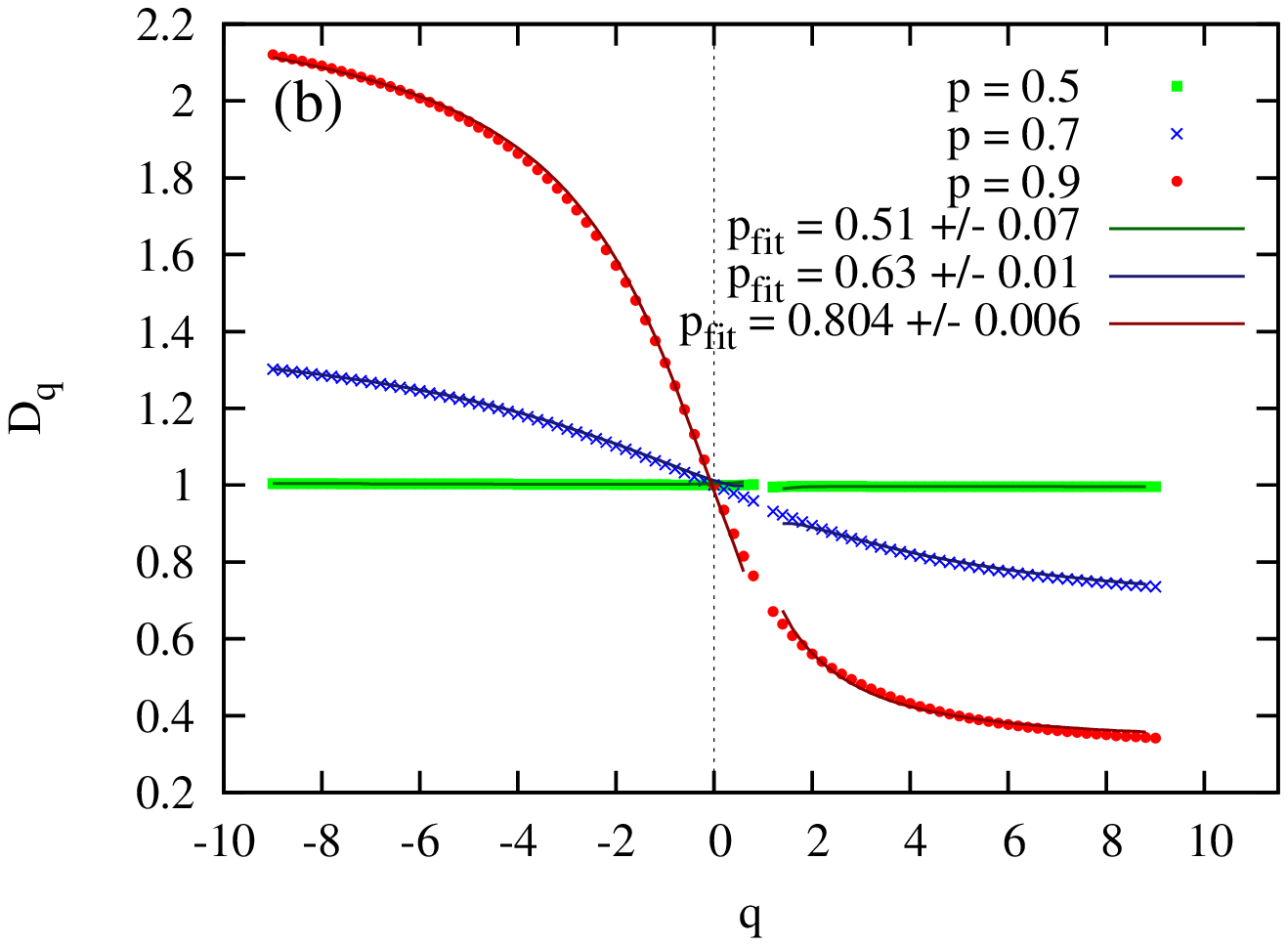}
\includegraphics[width=8cm]{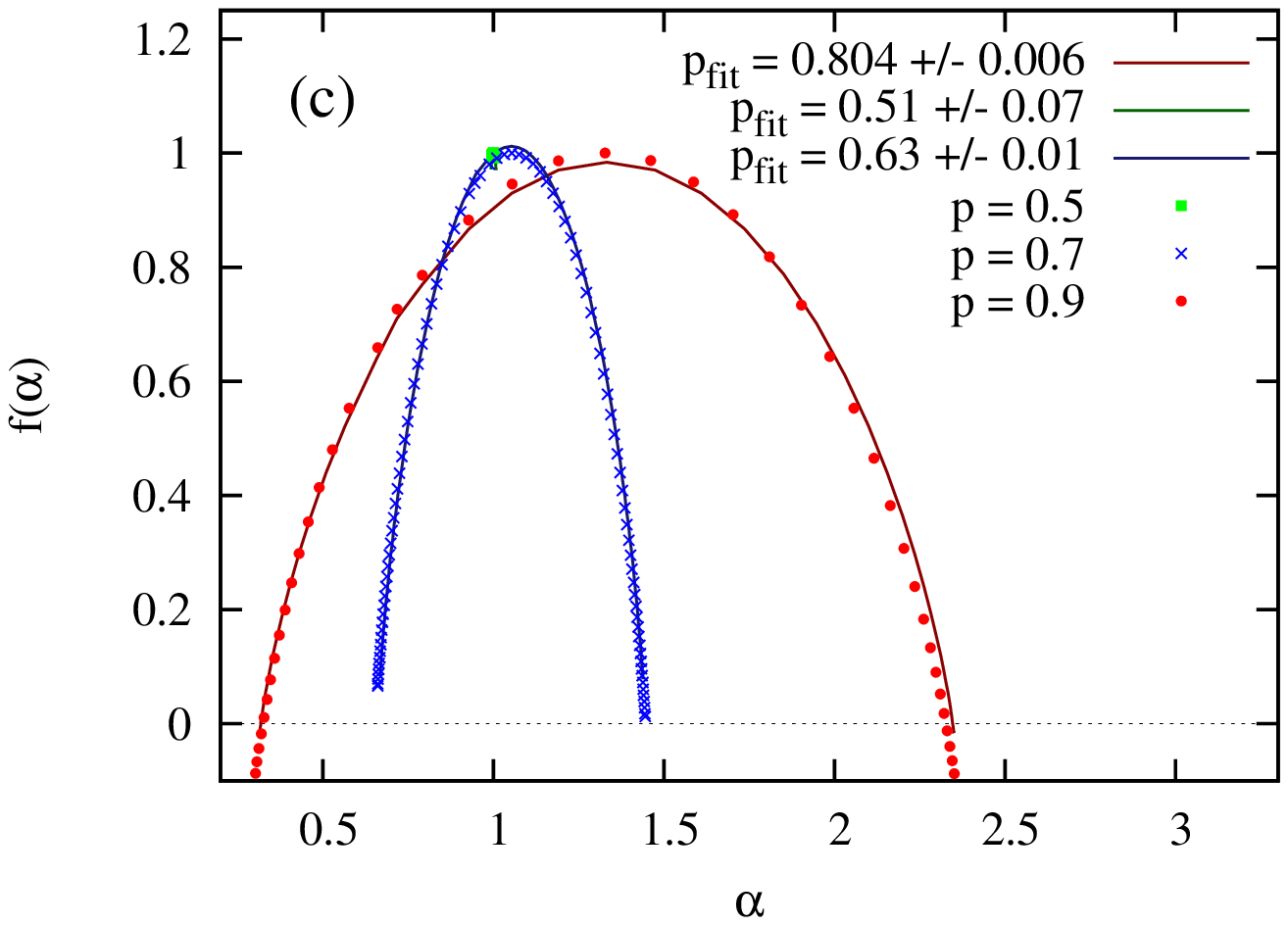}
\caption{The multifractality highlighted by the shape of R\'enyi scaling exponent $\tau_q$ vs $q$ (top-left panel ($a$)), the generalized dimension $D_q$ (top-right panel ($b$)), and the multifractal spectrum $f(\alpha)$ (bottom panel ($c$)). The case with $p=0.5$ is a monofractal, while the case with $p=0.7$ displays a multifractal degree smaller than for $p=0.9$. In the plots, solid lines represent the $p$-model fits performed on the scaling exponents $\tau_q$ and then transformed into the other quantities. The values of the fitting parameter $p_{fit}$ are indicated.}
\label{fig:tau_q_dq} 
\end{center}
\end{figure}

%
%
A different way of characterizing the intermittent behavior is the determination of the multifractal properties of the signal under study, in particular of generalized multifractal dimensions and the singularity spectrum associated with an appropriate measure~\cite{Paladin_Vulpiani_87}.
The mutifractal formalism~\cite{Frisch_Parisi_85,Halsey_86} was originally introduced in the context of fully developed turbulence and chaotic systems~\cite{Mandelbrot_82}, but later on it has become a standard tool to analyze phenomena observed in disordered system (see Ref.~\cite{Paladin_Vulpiani_87}). Multifractal analysis is able to capture the spatial disomogeneities of the turbulent energy cascade, so that {\it global} scale-invariance and self-similarity are usually associated to monofractal measures, while {\it local} scale-invariance, or local self-similarity, is associated a to multifractals.
For the analysis of our model fields, a suitable choice of an associated scalar quantity is the squared derivative along the trajectory $\partial_s$ of the velocity field component $v(s)$, estimated as the longitudinal velocity increment at the resolution scale, $\partial_s v(s)^2 = \Delta v(s,dr)^2$. 
To investigate the multifractal structure of this signal we use the standard box-counting method~\cite{vio_etal_92, Halsey_86}. Given the scalar signal $\Delta v(s)^2$, the generalized box-counting partition function of order $q$ is defined as
  \begin{equation}
    \chi_q(l)=\sum_{i=1}^{N(l)} \mu_i(l)^q  \,\,\, ,
    \label{eq:chi_q1}
  \end{equation}
where $N(l)$ is the minimum number of one-dimensional segments $Q_i(l)$ of length $l$ necessary to cover the trajectory $L$, and $\mu_i(l)$ is a suitably defined scale-dependent measure on the line:
  \begin{equation}
    \mu_i(l) = \frac{\sum_{s\in Q_i(l)} \Delta v(s)^2 }{\sum_{s\in L} \Delta v(s)^2}
    \label{eq:mu_i}
  \end{equation}
High values of $q$ in the partition function $\chi_q$ enhance the strongest singularities, say the most intense values of the signal under analisys, while small values of $q$ represent the regular regions. Conversely, negative values of $q$ emphasizes regions where the measure $\mu_i(l) $ is smaller, or the ``voids'' in the signal. 
The generalized dimensions $D_q$ are then formally defined by:
  \begin{equation}
    \label{D_q}
    D_q=\frac{1}{q-1} \lim_{l \rightarrow 0} \frac{\log \chi_q (l) }{\log l}
  \end{equation}
The definition given in equation~(\ref{D_q}) implies a scaling behavior of the partition function $\chi_q(l)$ for small $l$:
  \begin{equation}
    \label{chi_q}
    \chi_q(l) \sim l^{\tau_q} \; , \;\;\; \textrm{where} \;\;\; \tau_q = (q-1) D_q
  \end{equation}
and $\tau_q$ is the $q$-order ``mass'' exponent (also called R\'enyi scaling exponent) of the generalized partition function. The box-counting method consists of calculating the partition functions $\chi_q$, then derive $\tau_q$ from the power-law fit of $\chi_q$, obtain the generalized dimensions $D_q$ through equation~(\ref{chi_q}), and then the multifractal spectrum $f(\alpha)$ through a Legendre transform, given by:
  \[
  \begin{cases}
    f(\alpha) = q \; \alpha - \tau_q \\
    \alpha = \frac{d \tau_q}{dq}     \;\;\;\;\; . 
  \end{cases} 
  \]
\\
The latter basically gives the distribution of fractal dimensions of the subsets where the field has a given singularity strength~\cite{Chhabra_et_al_1989,Paladin_Vulpiani_87}.
Multifractal systems display nonlinear order dependence of the scaling exponents $\tau_q$, which implies non single-valued dimensions $D_q$, and which result in a broad multifractal spectrum $f(\alpha)$~\cite{Chhabra_et_al_1989,Paladin_Vulpiani_87}. 
In order to test the multifractality of our model as a signature of intermittency, we thus compute the partition functions $\chi_q(l)$ by varying the value of the exponent $q \in [-9,9]$ with step $dq = 0.2$, for each of the ten trajectories considered in the domain of the system, and for the three isotropic runs with $p = 0.5, 0.7, 0.9$. For each run, we then compute the average partition functions over the ten trajectories, as already done for the other statistical quantities, and we derive $\tau_q$ by fitting the functions $\chi_q(l)$ to power laws. Partition functions and the relative power-law fits are shown in Figure~\ref{fig:funz_partiz} for the run with $p=0.7$. 
The behavior of $\tau_q$ as function of $q$ is the result of this procedure, and is depicted in Figure~\ref{fig:tau_q_dq} (panel ($a$)). The linear dependence observed for the run with $p=0.5$ indicates fractal characteristics, while the degree of multifractality increases for larger $p$. This is also evident by looking at the generalized dimension $D_q$, shown in the right panel ($b$) of Figure~\ref{fig:tau_q_dq}, which is constant for $p = 0.5$ and increasingly broadens for larger $p$. The same behaviour is observed in the multifractal spectrum $f(\alpha)$, shown is the bottom panel ($c$) of Figure~\ref{fig:tau_q_dq}. In the non-intermittent case, the spectrum is single-valued, indicating that one single singularity exponent characterizes the whole space. As the model parameter $p$ is increased to induce intermittency, the spectrum becomes evidently broader, indicating a greater variety of the singularity exponents, or inhomogeneity of the cascade.

Finally, in order to have a more quantitative estimate of multifractal properties of the field, we fit the scaling exponents $\tau_q$ with the $p$-model prescription $\tau_q = - log_2 [p^q + (1-p)^q]$~\cite{meneveau87}. We then compare the values obtained from the fit, $p_{fit}$, with the prescribed intermittency parameter $p$, as already done for the structure functions analysis. The fits and the values of $p_{fit}$ are indicated in the three panels of Figure~\ref{fig:tau_q_dq}. The graphs show a good qualitative agreement, i.e. multifractality grows as the imposed intermittency increases. However, the quantitative comparison between $p$ and $p_{fit}$ shows some discrepancy, the fitted values being somewhat smaller than the imposed ones for the two intermittent runs. This could be due to the model limitations in capturing the finer geometrical properties of the intermittent structures. The specific choice of the field used for the analysis in this paper couls also have an effect on the measure. Different such choices have been tested giving similar results, but a more detailed study is deferred to a separate work. 
Nevertheless, the overall response of the model to multifractal analysis is satisfactory, at least qualitatively.
%
%
%
%

\subsection{Anisotropic turbulence}
When anisotropy is introduced in the model, it is necessary to test the intermittency as a function of the virtual trajectory direction. Since the imposed anisotropy is gyrotropic, it is sufficient to study the angular variation with respect to the anisotropy axis (in the present case along the $z$ axis), corresponding for example to the mean magnetic field direction in a MHD turbulence. The imposed symmetry also allows to use one quadrant only, so that ten trajectories have been selected to scan the non-gyrotropic angle $0^\circ<\theta<90^\circ$. Each of these trajectories has been divided in ten subsets of size $L\gg\ell_I$, and the results of the different diagnostic tools have been averaged for each angle. Again, their standard deviation represents the statistical uncertainty. The analysis has been performed on the non-intermittent case, i.e. $p=0.5$, and on the intermittent case with $p=0.7$.
\subsubsection{Spectral analysis.}
\begin{figure}
\begin{center}
\includegraphics[scale=1.]{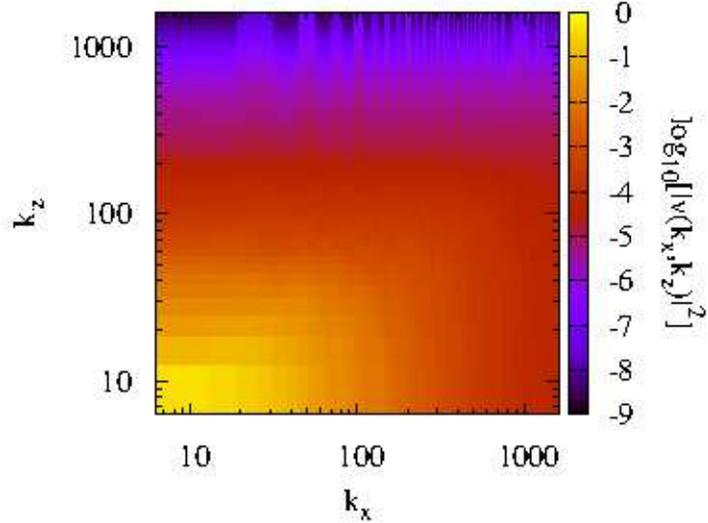}
\caption{Isocontours of the two-dimensional spectrum in the plane $k_x,k_y$ for the total power associated to the intermittent field $v$ in the anisotropic case. The image refers to the case $p=0.7$.}
\label{fig-spettro2d-anis}
\end{center}
\end{figure}

\begin{figure}
\begin{center}
\includegraphics[scale=0.5]{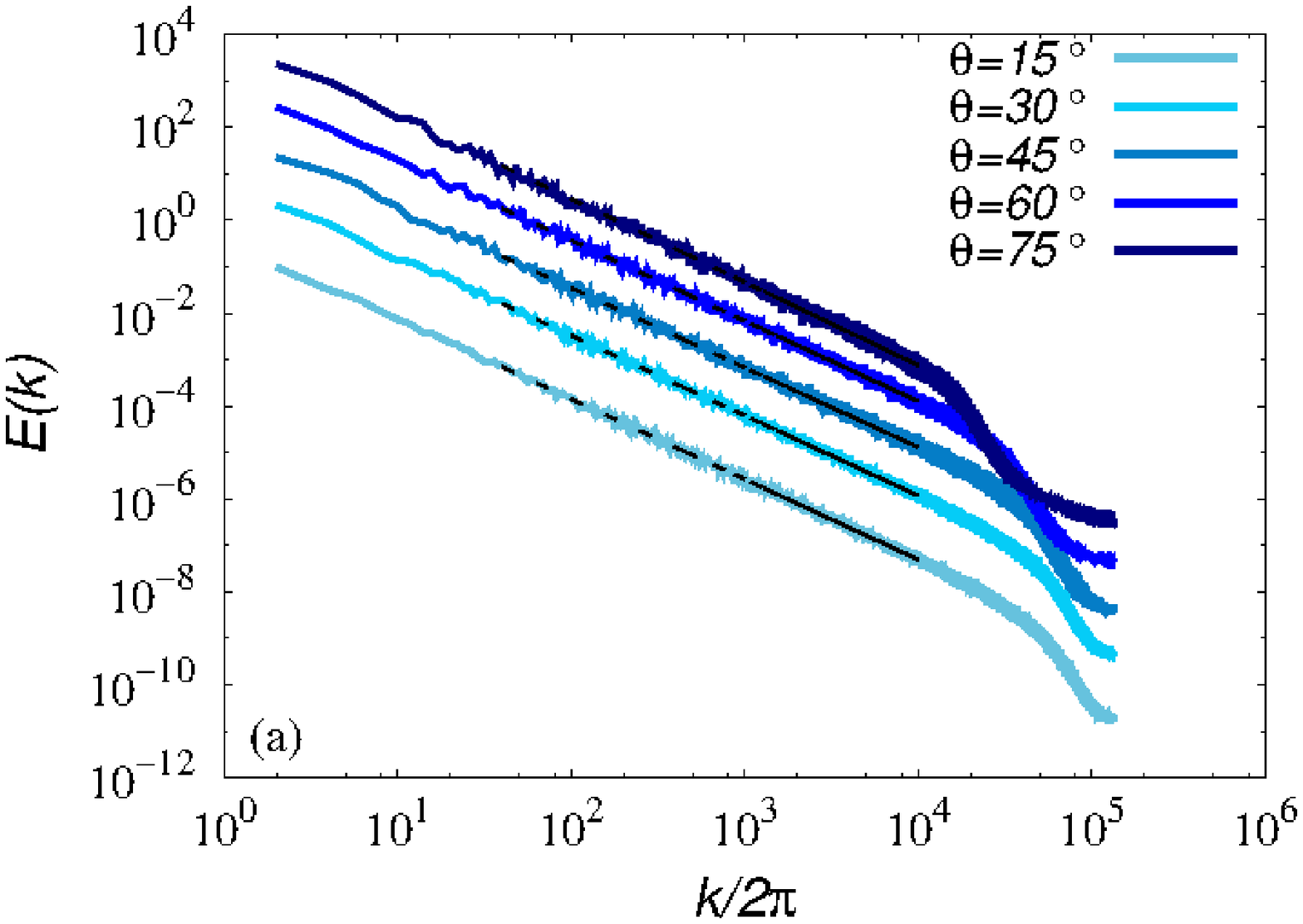} 
\includegraphics[scale=0.5]{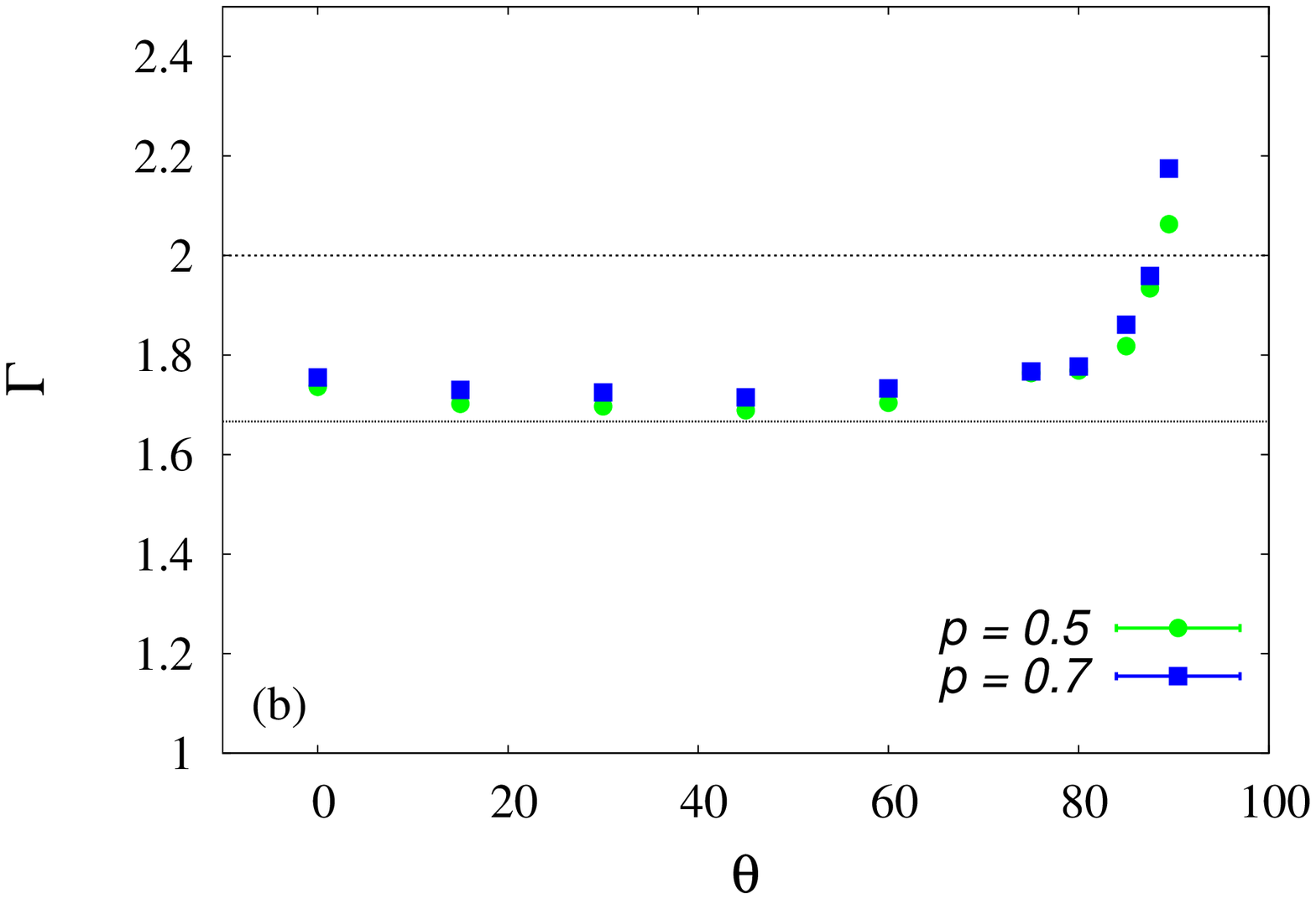}
\caption{Top panel ($a$): the one-dimensional power spectrum $E(k)$ of the virtual trajectories within the synthetic field, shown here for five different directions at an angle $\theta$ with respect to the anisotropy direction for the $p=0.7$ case. Power-law fits are also superposed, showing good agreement with the imposed Kolmogorov-like spectrum for intermediate angles. Bottom panel ($b$): the power-law index as a function of the virtual trajectory angle $\theta$, for $p=0.5$ and $p=0.7$. The deviation toward larger values for $\theta > 80^\circ$ is evident.}
\label{fig-spettri-anis}
\end{center}
\end{figure}

The two-dimensional spectrum for the $p=0.7$ run is shown in Figure~\ref{fig-spettro2d-anis}, where the anisotropic distribution of power is evident. 
%
%
Figure~\ref{fig-spettri-anis} shows the power spectra for different angles $\theta$ between $15^\circ$ and $75^\circ$, for the intermittent case $p=0.7$ (top panel ($a$)). The fitted power-law index as a function of the angle $\theta$ is shown in the bottom panel ($b$) of the same figure, both for the intermittent and for non-intermittent runs. As can be seen, the spectral index is reasonably constant for intermediate angles $15^\circ < \theta < 60^\circ$, and roughly coincides with the prescribed Kolmogorov value $\Gamma \simeq 5/3$. For quasi-perpendicular trajectories with $\theta > 80^\circ$, the spectral index increases, and reaches values as large as $\Gamma=2.1$. This behavior is qualitatively consistent with the prediction of critically balanced turbulence~\cite{goldreich95} and with some observations in numerical simulation and in solar wind measurements~\cite{cb-tim,cb-chen}. 
%
%
%
\subsubsection{Structure Functions.}
\begin{figure}
\begin{center}
\includegraphics[scale=0.5]{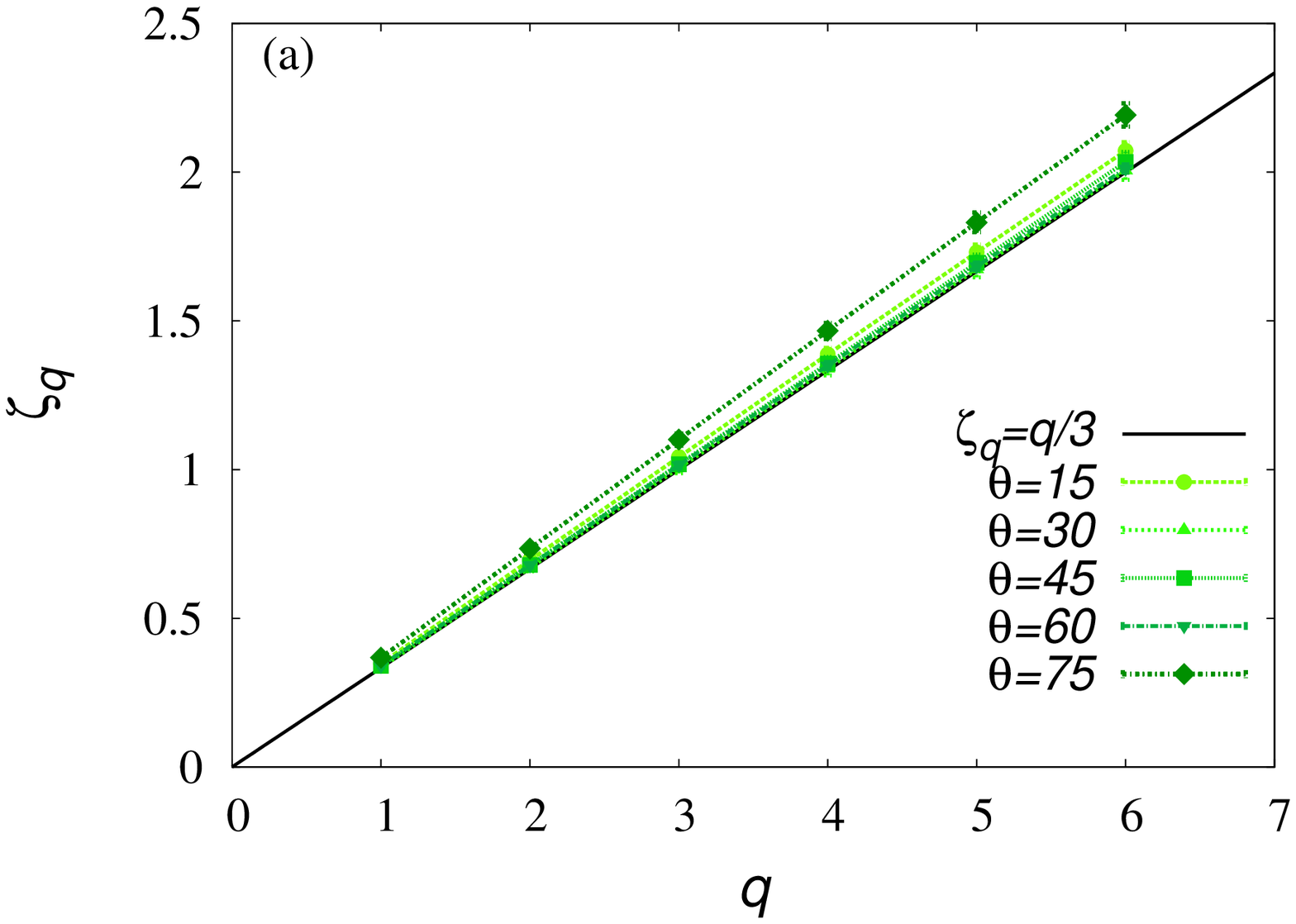} \\
\includegraphics[scale=0.5]{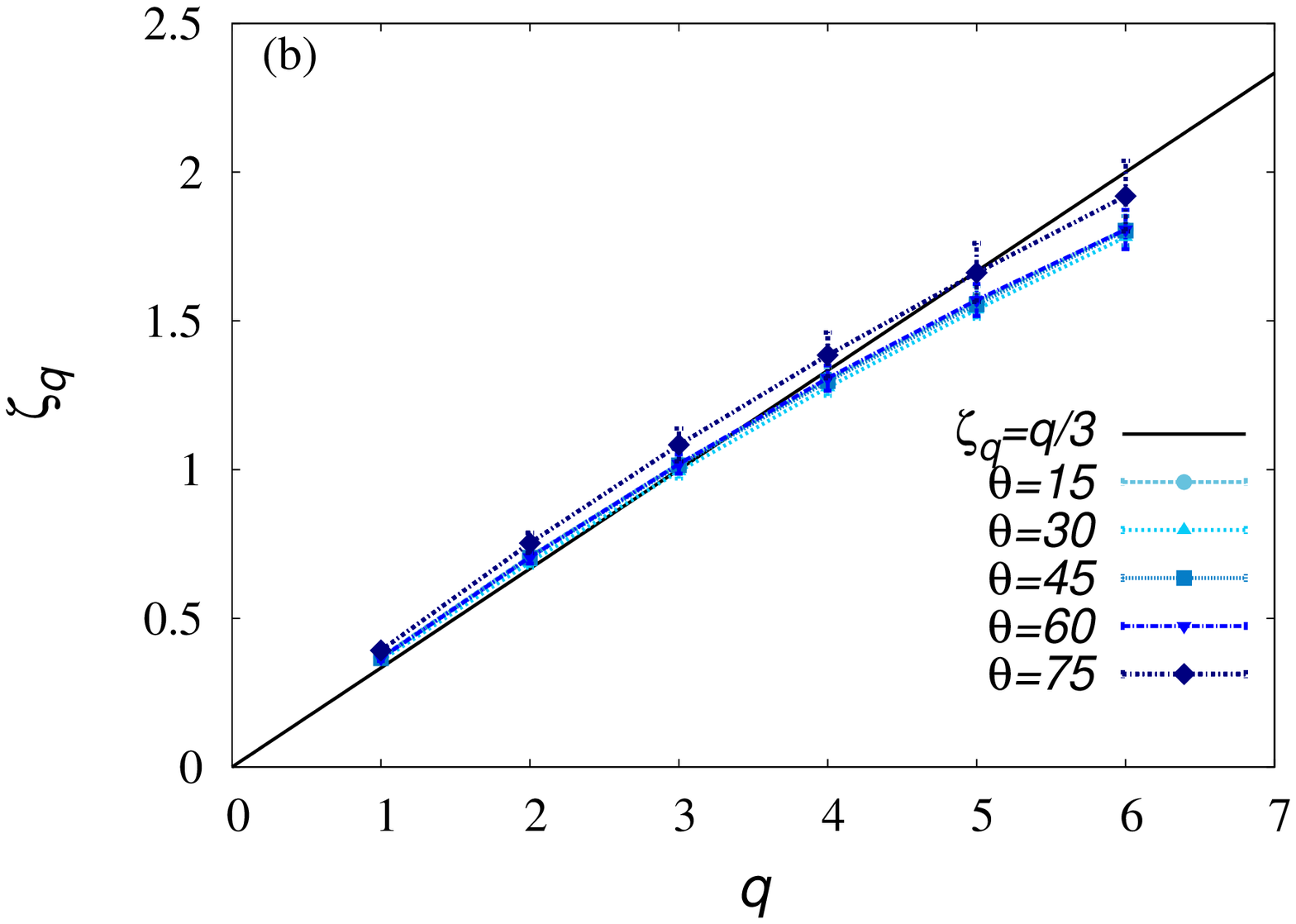}
\includegraphics[scale=0.5]{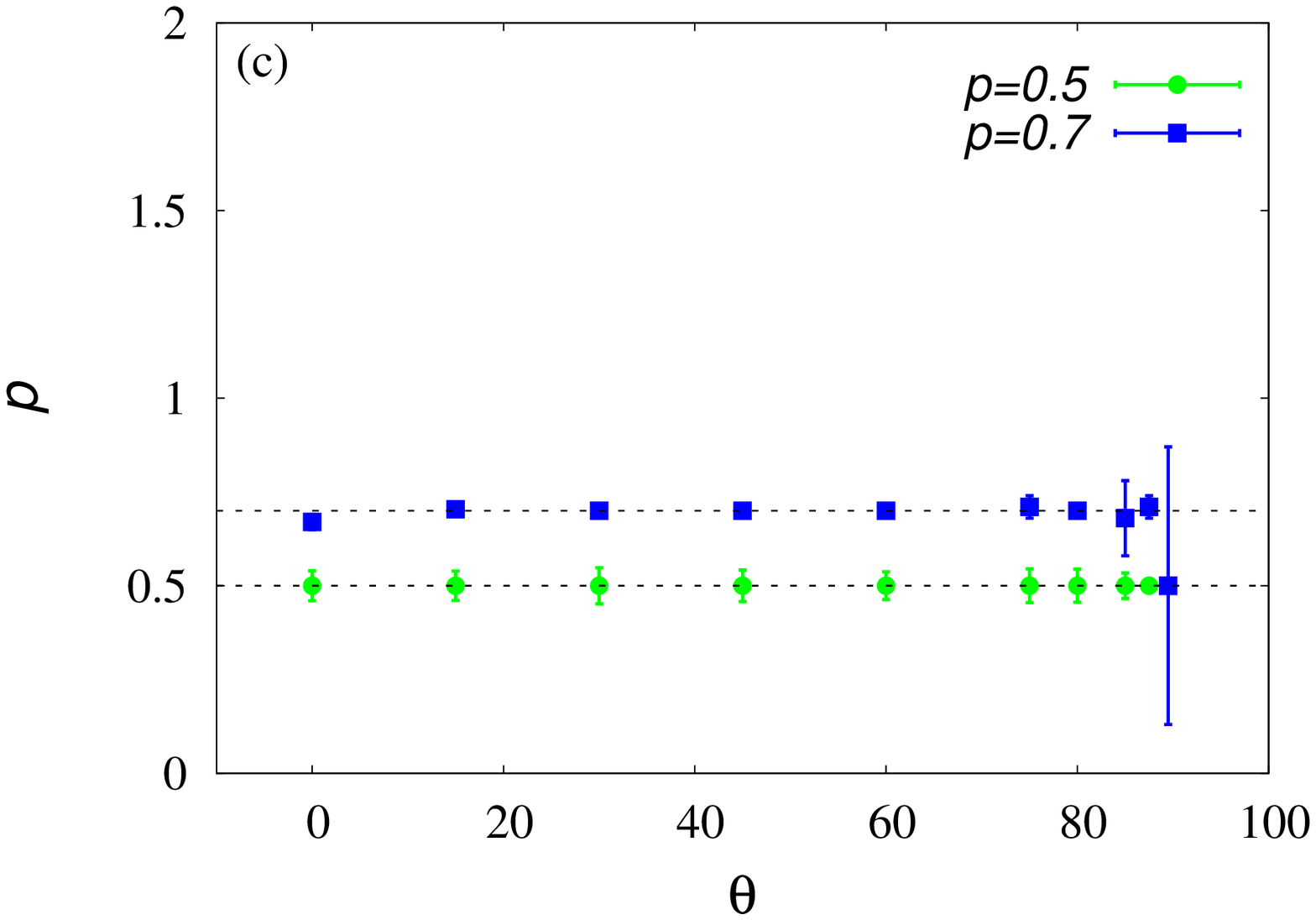}
\caption{The anomalous scaling of the structure functions for five different angles $\theta$ for the non-intermittent case $p=0.5$ (panel $a$) and for the intermittent case $p=0.7$ (panel $b$). $p$-model fits are shown as thick solid lines. Bottom panel ($c$): the angle dependence of the fitting parameter $p_{fit}$ for the two cases, with the horizontal lines indicating the input values $p$.}
\label{fig-strucfunc-anis}
\end{center}
\end{figure}

In order to evaluate the effects of anisotropy on intermittency, in Figure~\ref{fig-strucfunc-anis} we show the structure functions scaling exponents $\zeta_q$ for five different values of the angle $\theta$, for the two runs with and without intermittency (top ($a$) and central ($b$) panel). As for the isotropic case, the fit of the scaling exponents with the $p$-model provides a quantitative estimate of intermittency through the parameter $p$, which is plotted in the bottom panel ($c$) of Figure~\ref{fig-strucfunc-anis} as a function of the angle $\theta$, for the intermittent run considered in this Section (the non-intermittent case consistently provides $p=0.5$). It is evident that even in the presence of anisotropy, the intermittency prescription is recovered in the synthetic data (see Ref.s~\citep{anis06,anis10,anis15,pei} for recent results on intermittency in solar wind anisotropic turbulence). Only the case at $\theta=90^\circ$ displays a discrepancy, showing no intermittency even when $p=0.7$. This is probably due to the shape of the synthetic eddies along the axes, which is also responsible for the weak anisotropy of the spectral power in the isotropic case. Once again, this suggests that for an optimal response of the model, trajectories should be selected with an (even small) angle with respect to the system axes.
%
%
%
\subsubsection{Kurtosis.}
\begin{figure}
\begin{center}
\hskip 24pt \includegraphics[scale=0.5]{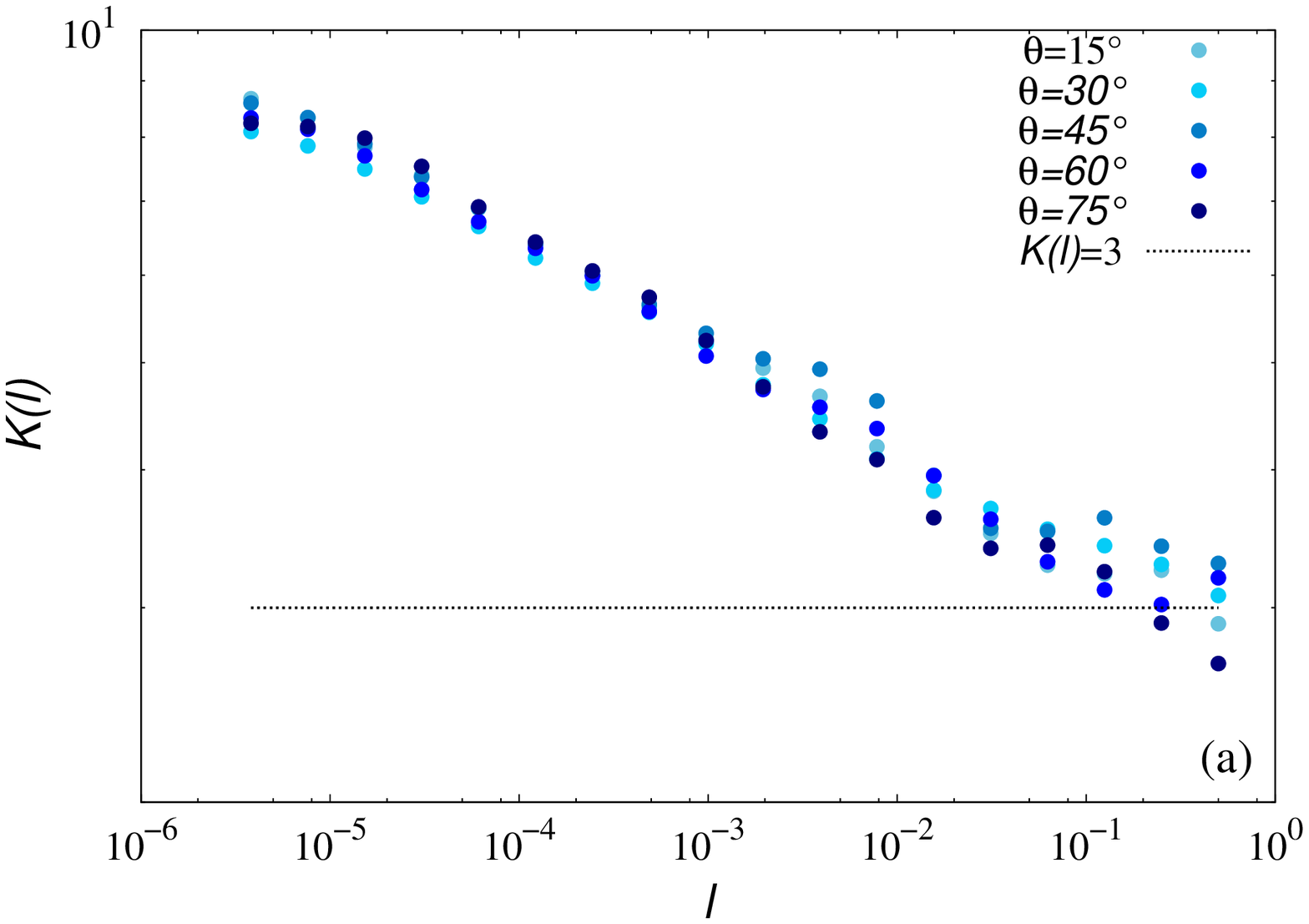}
\includegraphics[scale=0.55]{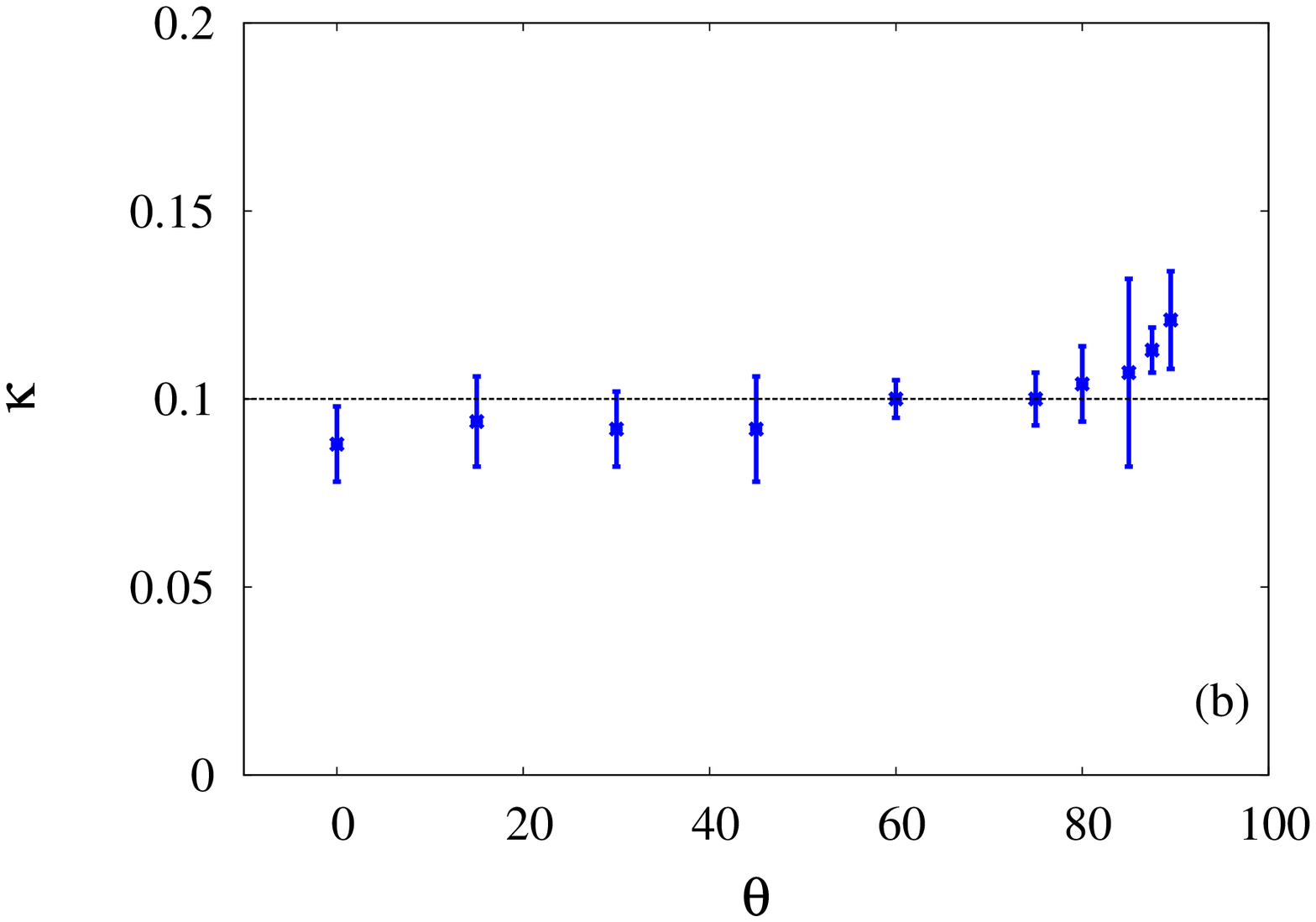}
\caption{Top panel ($a$): The scaling dependence of the Kurtosis $K$ for the $p=0.7$ case, and for five angles $\theta$. The Gaussian value $K=3$ is indicated. Bottom panel ($b$): The angular dependence of the scaling exponent $\kappa$, which shows an increase at large angles, as observed for the spectral index $\Gamma$ (Figure~\ref{fig-spettri-anis}).}
\label{fig-flat-anis}
\end{center}
\end{figure}

Finally, in Figure~\ref{fig-flat-anis} we show the variation of the kurtosis with the angle, for the intermittent case (top panel ($a$)); as expected, the non-intermittent case gives Gaussian values $F=3$ and $\kappa=0$ at all angles (not shown). When intermittency is present, the overall effect of anisotropy is to modulate the scaling exponent $\kappa$ of the kurtosis in response to the variations of the spectral exponent $\Gamma$ increase with the angle (see Figure~\ref{fig-spettri-anis}), and in particular for large angles $\theta > 80^\circ$ (bottom panel ($b$)). 
%
%
The anisotropic realization of the synthetic turbulence presented here is therefore able to capture the major characteristics of spectral anisotropy, and to preserve the intermittency properties.


\section{Conclusions}
\label{Section:conclusions}%

Synthetic turbulence models represent a useful tool which can be used in a variety of situations, mainly when it is necessary to have a realistic representation of a turbulence (either hydrodynamic or MHD) with an extended inertial range. This happens typically in  astrophysical contexts, like in the solar wind, where in-situ measurements have shown the presence of a turbulence with a spectrum extending over several decades of spatial scales.

In this paper we have presented and discussed a new model of synthetic turbulence, belonging to the class of ``wavelet-based" models, in which the synthetic field is obtained by a superposition of base functions at different spatial scales, whose amplitude is determined so as to reproduce a given spectral law for the turbulent field. Moreover, the model reproduces intermittency in the turbulent field by means of a $p$-model technique \cite{meneveau87}, in which the spectral energy flux from a given spatial scale to the smaller one is unevenly distributed in space. The modelled turbulent field is three-dimensional in space and solenoidal, so it can be used to describe either an incompressible flow or a turbulent magnetic field. No time dependence is included in the model.

Our model shares many aspects with models by Juneja et al.~\cite{juneja94} and by Cametti et al.~\cite{cametti98}, but with relevant differences in the algorithm. In fact, one important limitation in the 3D model by Cametti et al.~\cite{cametti98} is in the memory requirement, which rapidly increases when considering increasing spectral width. In the study presented by these authors the spectral extension is limited to (about) two decades. The algorithm employed by our model has been designed so as to avoid both large memory employments and long computational times in the evaluation of the turbulent field at a given spatial position. In particular, the computational time $t_C$ scales proportional to $\log_2 (L_0/\ell_{N_s})$, where $\ell_{N_s}$ is the smallest scale included in the model. This is perhaps the most important feature of the model, because it allows to describe a turbulence with a very extended spectral range using a modest computational effort. All the results presented in this paper have been obtained running the model on a desktop computer: in a typical run, which took about 20 min of CPU time, the turbulent field with a spectral extension between 4 and 5 decades has been calculated in a number of spatial positions of the order of $2.5 \times 10^6$. Moreover, the memory requirement is very low: each time the field is to be evaluated at a given position, all the parameters defining the involved eddies are re-calculated without keeping any information in the computer memory.

The model contains few parameters, namely:
(i) the parameter $h$, which contributes to determine the index $\Gamma$ of the power-law spectrum; 
(ii) the parameter $p$, which sets the ``level" of intermittency and contributes (to a smaller extent) to determine $\Gamma$; 
(iii) the spectral width, fixed by the ratio $L_0/\ell_{N_s}$. 
Such parameters can be tuned in order to reproduce different physical situations. 
Finally, we explored the possibility to include an anisotropic spectrum, trying to reproduce the situation described by the so-called ``critical balance" principle, postulated by Goldreich \& Sridhar~\cite{goldreich95} in the case of a MHD turbulence, often advocated for the description of solar wind turbulence.

In order to assess the validity of the model and its reliability in reproducing realistic flows, we have run the standard diagnostics for intermittent turbulence and verified that the synthetic field indeed possesses the characteristics that were chosen as input. To this aim, we have obtained a series of isotropic runs by fixing the scaling exponent $h$, and varying the intermittency parameter, which was given three values: $p=0.5$ (no intermittency), $p=0.7$ (standard Navier-Stokes intermittency), $p=0.9$ (strong intermittency). We have then extracted synthetic one-dimensional cuts within the model domain, and have applied time-series analysis techniques: autocorrelation function, power spectrum, probability distribution functions of the field increments, their structure functions, the kurtosis, and a standard multifractal analysis. All the tests gave satisfactory results, showing that the synthetic data reproduce well the required conditions of spectral scaling and intermittency. A small anisotropy originated by the particular shape of the eddy functions is present along the three axes of the system. This was easily mediated by choosing trajectories with an angle with the three axes. 
We have also explored the geometry of the system by using two anisotropic runs, with $p=0.5$ and $p=0.7$, and by imposing the critical balance conditions. Even in the anisotropic case, the output satisfactorily reproduces the expected values of spectral slope and intermittency for all the observables. We can conclude that the model provides a good representation of intermittent turbulence, and is sensitive to the choice of the input parameters, which allows to fine tune the type of turbulence as desired. 

It is important to acknowledge that the present version of our model is not able to reproduce the skewness of the field increments PDFs, i.e. their non-vanishing third-order moment, universally observed in fully developed turbulence. An improved version of the model that accounts for the appropriate description of the skewness is currently in progress. 

Finally, we wish to note that a preliminary version of the present model has been recently employed to study the problem of energetic particle diffusion in a magnetic turbulence~\cite{pucci16}.  The highly suprathermal speed of the energetic test particles, as observed for example in the solar wind, allowed the use of the static turbulent field generated by our model.
That investigation has singled out relevant effects on the particle transport related to both large spectral extensions and to intermittency. Thus, a representation of a 3D turbulence with a wide spectrum, as well as a tunable level of intermittency, have been crucial aspects of employing the present synthetic turbulence model in this study. Furthermore, when using our model to run test-particle simulations the integration of particle trajectories is considerably simplified by the possibility to calculate the turbulent field directly at any spatial position, thus avoiding interpolations on a spatial grid.

\begin{acknowledgments}
The authors wish to thank Fabio Lepreti for his valuable help in the redaction of this work. 
\end{acknowledgments}

\bibliography{basename of .bib file}

\end{document}